\documentclass[numberedappendix,referee]{aastex}

\usepackage{graphicx}
\usepackage{natbib}
\usepackage{ulem}
\usepackage{txfonts}

\slugcomment{Submitted to the Astrophysical Journal}

\shorttitle{}
\shortauthors{Mart\' inez Gonz\'alez et al.}

\begin{document}

\title{Unnoticed magnetic field oscillations in the very quiet Sun revealed by Sunrise/IMaX}
\author{M. J. Mart\' inez Gonz\'alez, A. Asensio Ramos, R. Manso Sainz, E. Khomenko, V. Mart\' inez Pillet}
\affil{Instituto de Astrof\'{\i}sica de Canarias, C/V\'{\i}a L\'actea s/n, 38200
La Laguna, Tenerife, Spain\\
Departamento de Astrof\'{\i}sica, Univ. de La Laguna, 38205, La Laguna, Tenerife, Spain}

\and

\author{S. K. Solanki}
\affil{Max-Planck-Institut f\"ur Sonnensystemforschung, 37191, Katlenburg-Lindau, Germany\\
School of Space Research, Kyung Hee University, Yongin, Gyeonggi, 446-701 Korea}

\and

\author{A. L\'opez Ariste}
\affil{THEMIS - CNRS UPS 853, C/ V\' ia L\'actea s/n. 38205, La Laguna, Tenerife, Spain}

\and

\author{Schmidt, W.}
\affil{Kiepenheuer-Institut f\"ur Sonnenphysik, Sch\"oneckstr. 6, 79104 Freiburg, Germany}

\and

\author{Barthol, P., Gandorfer, A.}
\affil{Max-Planck-Institut f\"ur Sonnensystemforschung, 37191, Katlenburg-Lindau, Germany}

\begin{abstract}
We present observational evidence for oscillations of magnetic flux density in the quiet areas of the Sun. The majority of magnetic fields on the solar surface have strengths of the order of or lower than the equipartition field (300-500 G). This results in a myriad of magnetic fields whose evolution is largely determined by the turbulent plasma motions. When granules evolve they squash the magnetic field lines together or pull them apart. Here we report on the periodic deformation of the shapes of features in circular polarization observed at high resolution with Sunrise. In particular, we note that the area of patches with constant magnetic flux oscillates with time, which implies that the apparent magnetic field intensity oscillates in antiphase. The periods associated to this oscillatory pattern is compatible with the granular life-time and change abruptly, which suggests that these oscillations might not correspond to characteristic oscillatory modes of magnetic structures, but to the forcing by granular motions. In one particular case, we find three patches around the same granule oscillating in phase, which means that the spatial coherence of these oscillations can reach 1600 km. Interestingly, the same kind of oscillatory phenomenon is found also in the upper photosphere.

\end{abstract}

\section{The magnetically, dynamically active quiet Sun}

As we enhance the spatial resolution and sensitivity of spectro-polarimetric measurements, it becomes increasingly evident that magnetic fields permeate the whole quiet Sun \citep{david_07, marian_08, lites_08, danilovic_10}. The observed Zeeman polarization signals in the quietest areas of the Sun are highly dynamic, having timescales compatible with the granulation \citep{lin_99, harvey_07, danilovic_10}. These weak magnetic flux concentrations emerge somewhat preferentially in granules, where plasma motions are more favorable for the magnetic fields to rise across the solar atmosphere \citep{marian_07, rebe_07, david_08, marian_09, gomory_10, danilovic_10}. In many of these features the field is organized in the form of $\Omega$-shaped loops and survives convective motions, the footpoints of these loops being passively advected by granular motions \citep{marian_09, rafa_10}. Importantly, some of these small-scale loops do reach higher layers, at least the lower chromosphere, hence linking the very quiet photosphere to chromospheric magnetism \citep{marian_09, marian_10, wiegelmann_10}.

In this letter we present another interesting feature of the small-scale solar magnetism. We find that both the area and the magnetic flux density of circular polarization patches displays quasi-periodic oscillations at constant magnetic flux.

Oscillations of line-of-sight velocity and brightness of small-scale magnetic features in solar network, faculae and plage areas have been detected by e.g. \cite{Lites+Rutten+Kalkofen1993, Krijer+etal2001, DeMoortel+etal2002, DePontieu+etal2003, Centeno+etal2009, Bloomfield+etal2006, Veccio+etal2009, jess_09}. However, a clear detection of magnetic field oscillations remains ellusive. Recently, \cite{fujimura_09} used HINODE spectro-polarimetric data with a spatial resolution of 0.32$''$, showing evidence of magnetic flux oscillations, interpreted by these authors as an indication for magneto-acoustic "sausage" and "kink" modes in individual elements of plages and pores. Very little (if anything) is known about possible oscillations of magnetic features in solar internetwork areas, with much less magnetic flux. Due to the usually low amplitude of the polarization signals, the observations necessary to study the temporal behavior of small-scale patches in the internetwork are rather demanding. Here we make use of the superb quality of the Sunrise/IMaX data and show that magnetic fields outside active regions also oscillate, representing the first evidence for magnetic field oscillations in the quiet Sun internetwork areas.

\section{The IMaX data}

We analyze disk center quiet Sun observations obtained with the IMaX instrument \citep{valentin_10} onboard the SUNRISE balloon borne observatory \citep{barthol_10, sami_10}. The IMaX instrument is a Fabry-P\'erot interferometer with polarimetric capabilities at the Fe\,{\sc i} line at 5250.2 {\AA}. The observational sequence consists of five filtergrams taken at $\pm$40, $\pm$80 and +227 m{\AA} from the Fe\,{\sc i} 5250.2 {\AA} line center. The field of view is 46.8$''\times 46.8''$, with a cutoff frequency of (0.15$''$-0.18$''$)$^{-1}$, the pixel size being 0.05$''$. We analyze two time series of about 23 and 40 min duration with a cadence of about 32 s and a noise level of $\sigma= 10^{-3}$ and $7\times 10^{-4}$ I$_\mathrm{c}$ in the circular and linear polarization (I$_\mathrm{c}$ being the continuum intensity).

We analyze the IMaX data before the phase-diversity reconstruction process, due to their lower noise level. The Fe\,{\sc i} line at 5250.2 \AA\ is a forbidden but resonant line and thus is extremely sensitive to the local temperature of the plasma \citep[e.g.][]{stenflo_solanki_87, lagg_10}. We use the flux density ($\phi$ in Mx cm$^2$) derived from the weak field approximation instead of the directly observed Stokes V signal. The weak field approximation states that the Stokes V profile is proportional to the derivative of the intensity profile and to the magnetic flux density. Taking into account the simultaneous information contained in the intensity and the Stokes V profile, we can decouple the thermodynamical and the magnetic effects. The magnetic flux density is estimated following a least-squares minimization \citep[see][]{marian_11}. It can be expressed as:
\begin{equation}
\phi=-\frac{\sum_i \frac{\partial I}{\partial \lambda}_i V_i}{C\sum_i
(\frac{\partial I}{\partial \lambda}_i)^2}
\end{equation}
which uses all the wavelength samples across the profile (index $i$)
and is therefore more accurate than determinations based on single
magnetogram measurements. The constant $C$ is defined as $4.6686\times
10^{-13}\lambda_0^2 \bar{g}$, $\lambda_0$ being the central wavelength
(in {\AA}) and $\bar{g}$ the effective Land\'e factor of the transition. The derivatives of the intensity profile have been computed numerically using the observed data. For this kind of IMaX data, this approximation is valid for magnetic fields up to \hbox{$\sim 1$ kG}. The estimated error is \hbox{$\sigma_B=4$ Mx cm$^{-2}$}.
If we assume that the magnetic fields are resolved (filling factor is one), Eq. 1 gives the longitudinal magnetic field strength in G.

\begin{figure*}[!t]
\centering{
\includegraphics[width=0.16\textwidth, bb= 115 399 417 701]{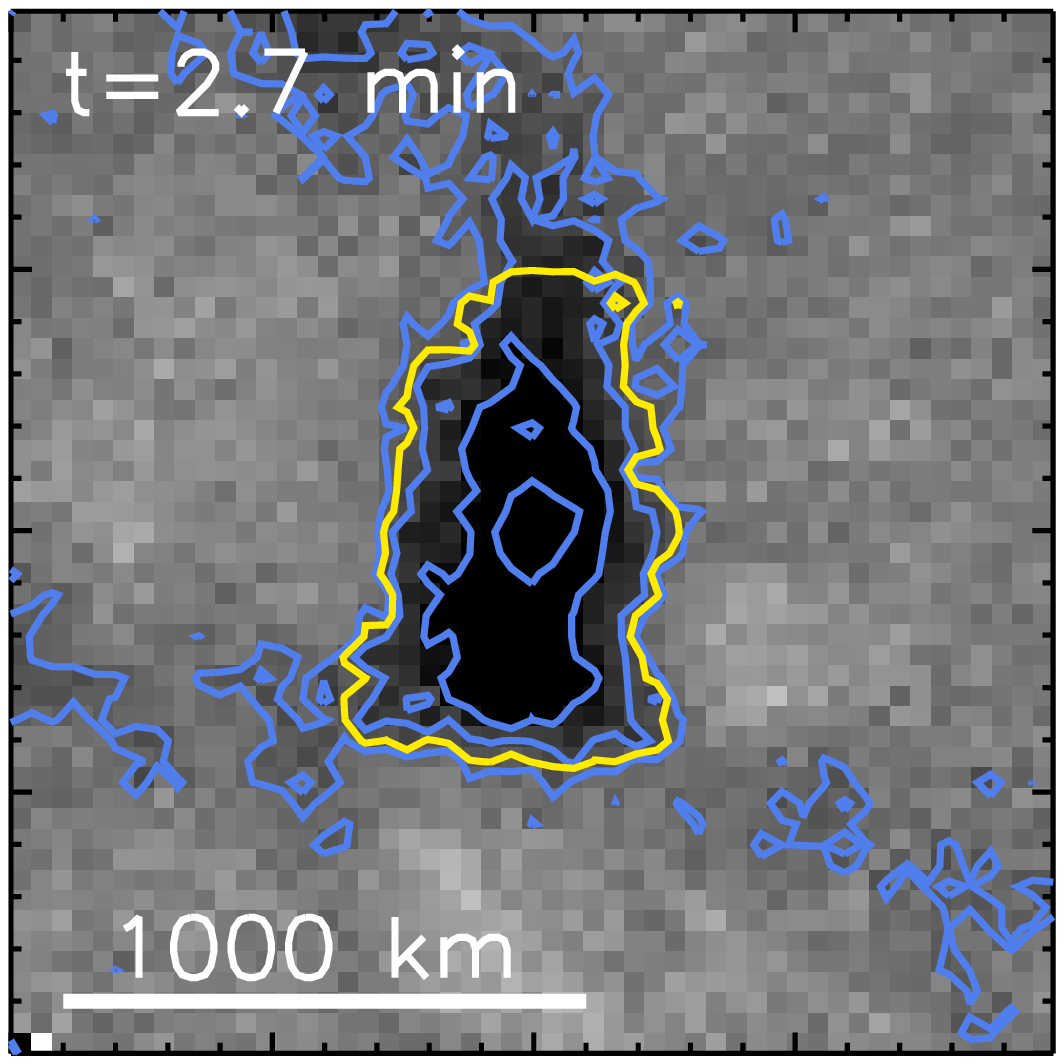}
\hspace{-0.15cm}
\includegraphics[width=0.16\textwidth, bb= 115 399 417 701]{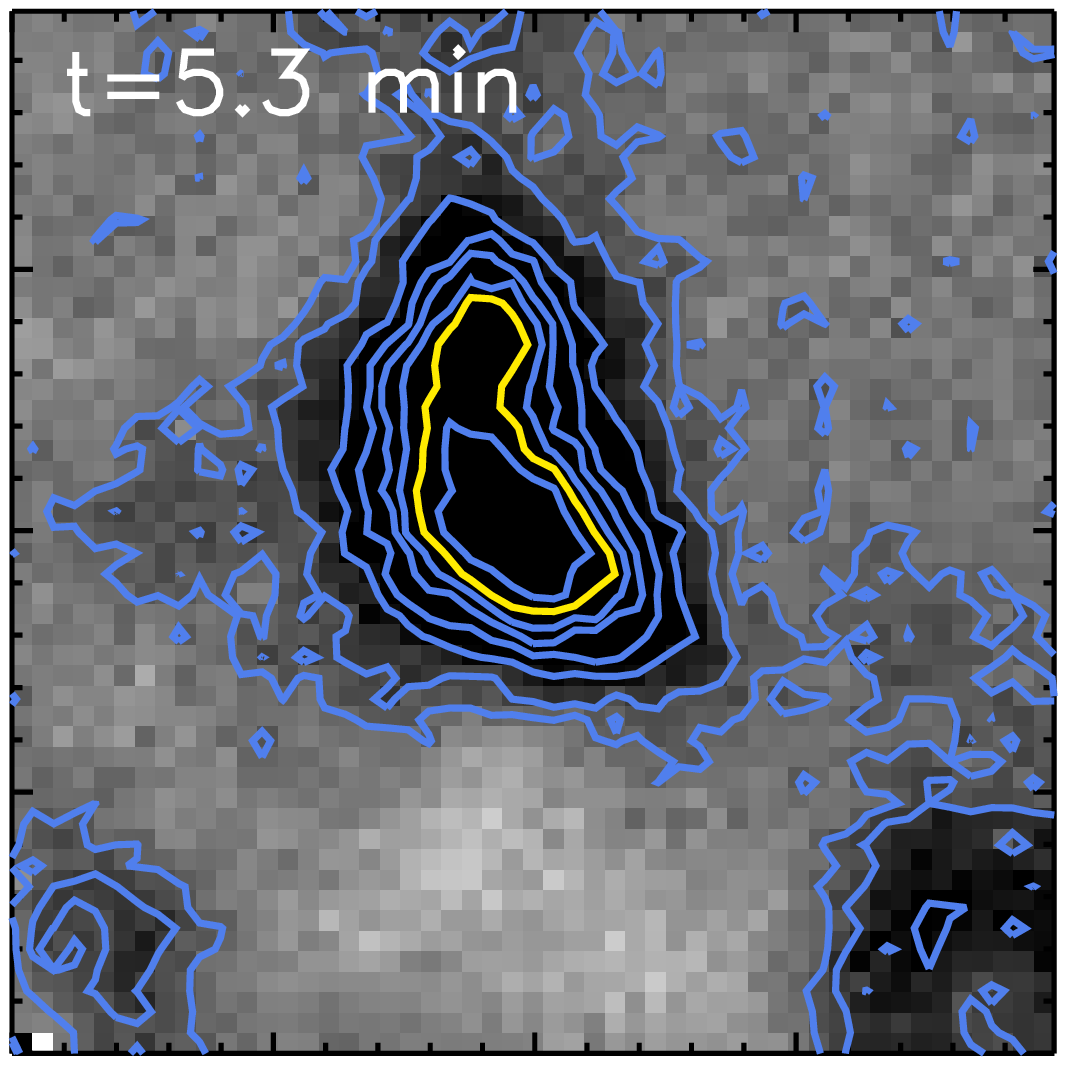}
\hspace{-0.15cm}
\includegraphics[width=0.16\textwidth, bb= 115 399 417 701]{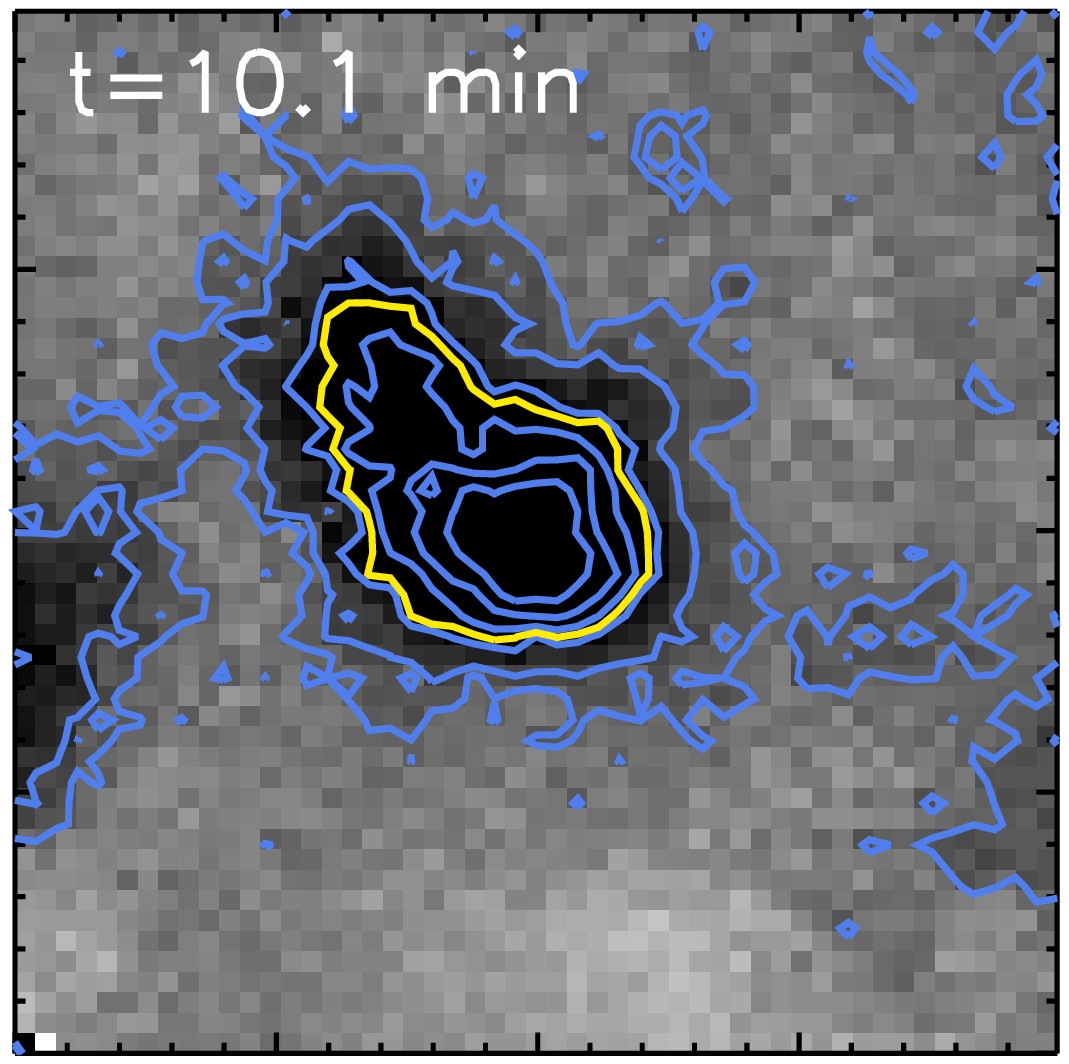}
\hspace{-0.15cm}
\includegraphics[width=0.16\textwidth, bb= 115 399 417 701]{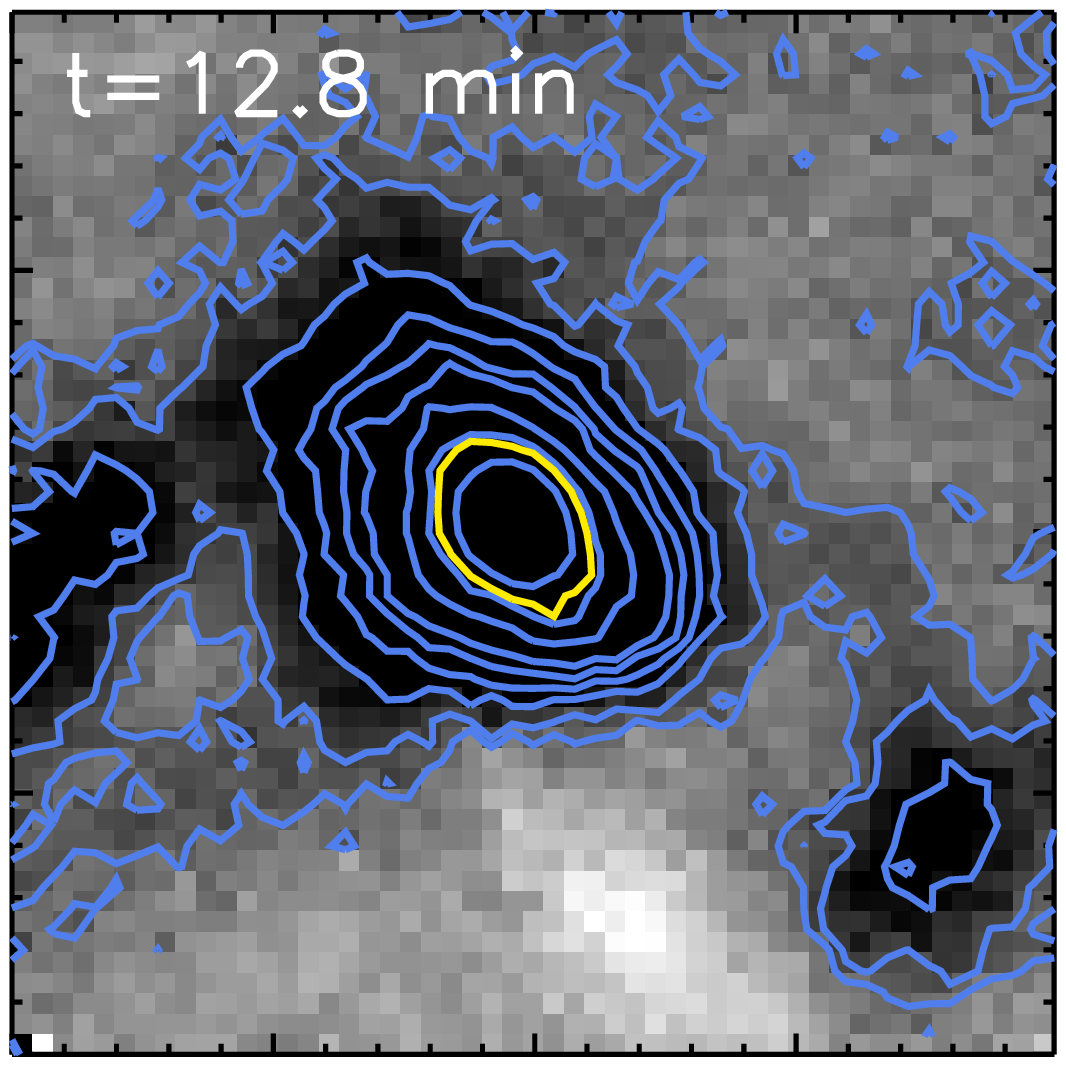}
\hspace{-0.15cm}
\includegraphics[width=0.16\textwidth, bb= 115 399 417 701]{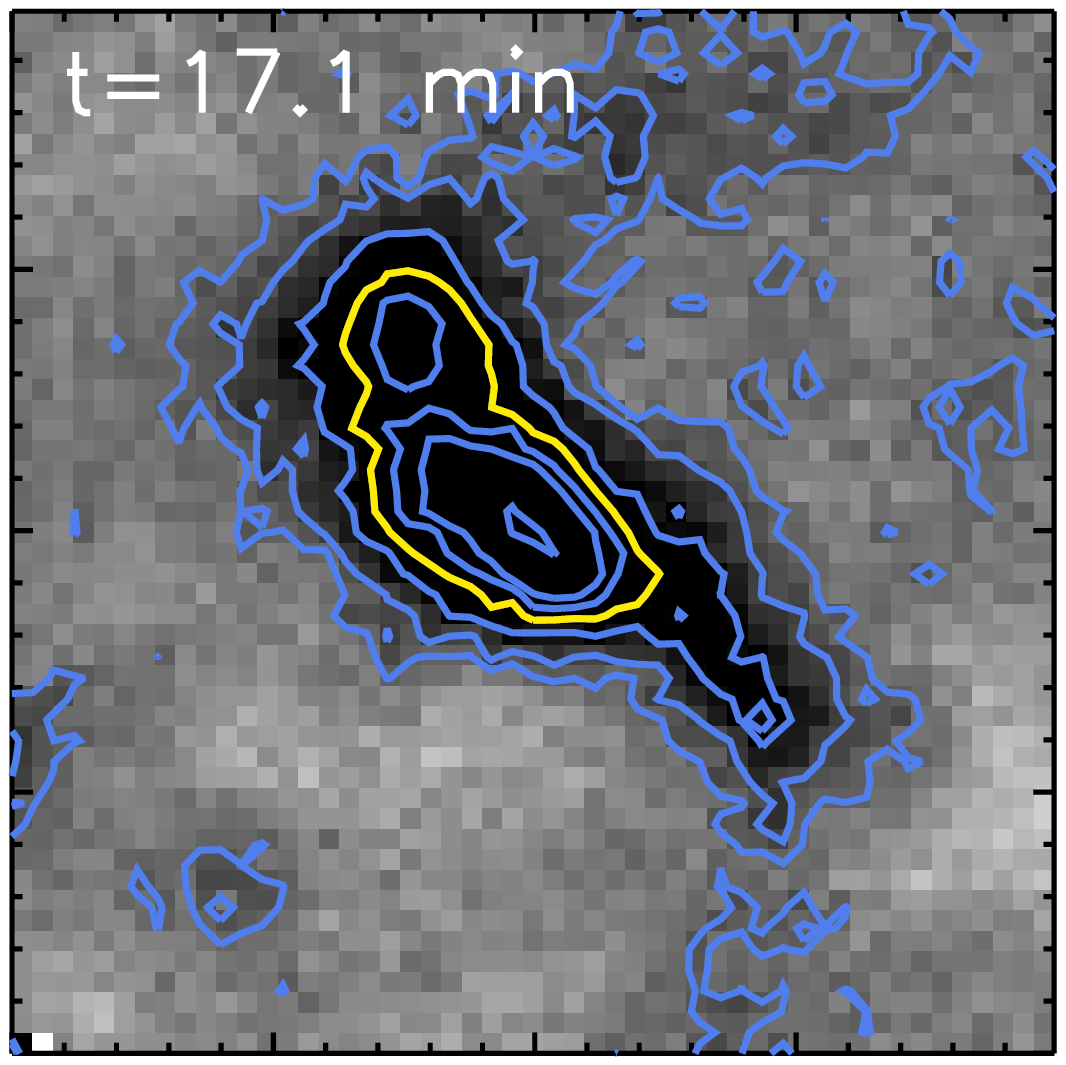}
\hspace{-0.15cm}
\includegraphics[width=0.16\textwidth, bb= 115 399 417 701]{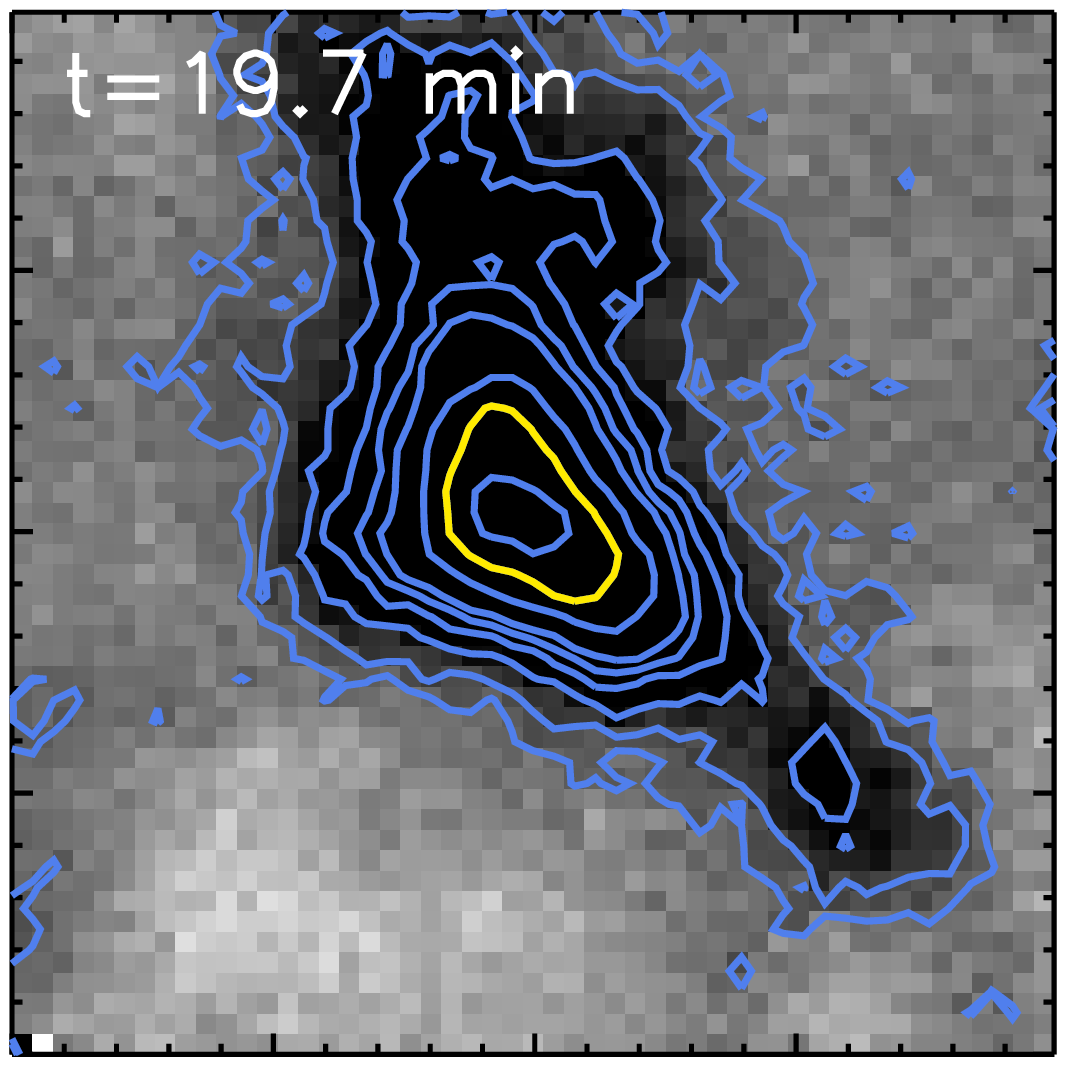}}
\caption{Time evolution of a weak circular polarization patch (corresponding to the top right panel of Fig. \ref{var_area}). The black and white background represents the magnetic flux density computed in the weak field approximation, saturated to $\pm 20$ Mx cm$^{-2}$. Blue lines represent iso-magnetic flux densities of -130,-100,-75,-50,-40,-30,-20,-10, and -5 Mx cm$^{-2}$. Yellow line is the iso-magnetic flux density contour containing a time-constant magnetic flux of -4.5$\times 10^{16}$ Mx.}
\label{ej1}
\end{figure*}

\begin{figure*}[!t]
\centering{
\includegraphics[width=0.16\textwidth, bb= 116 400 418 702]{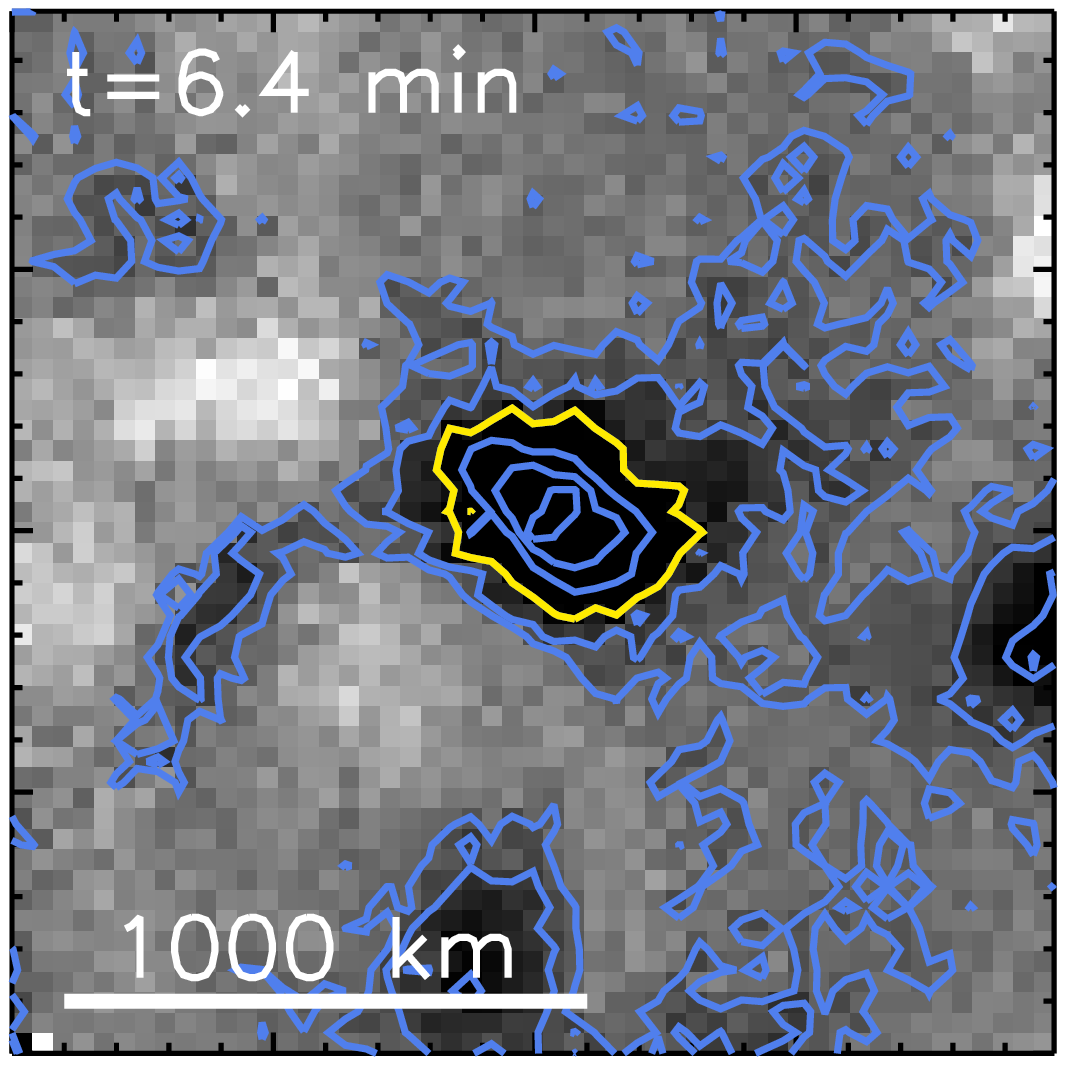}
\hspace{-0.15cm}
\includegraphics[width=0.16\textwidth, bb= 116 400 418 702]{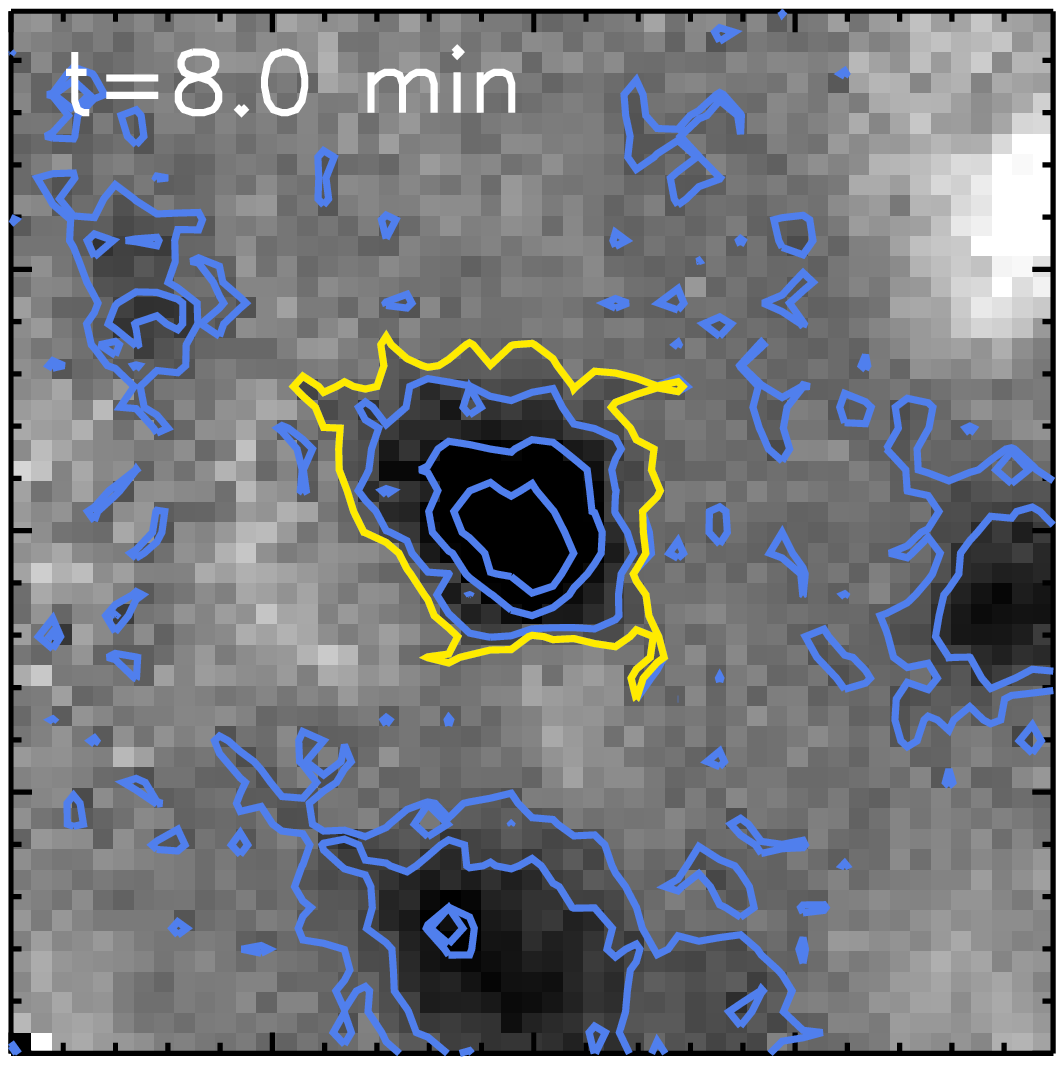}
\hspace{-0.15cm}
\includegraphics[width=0.16\textwidth, bb= 116 400 418 702]{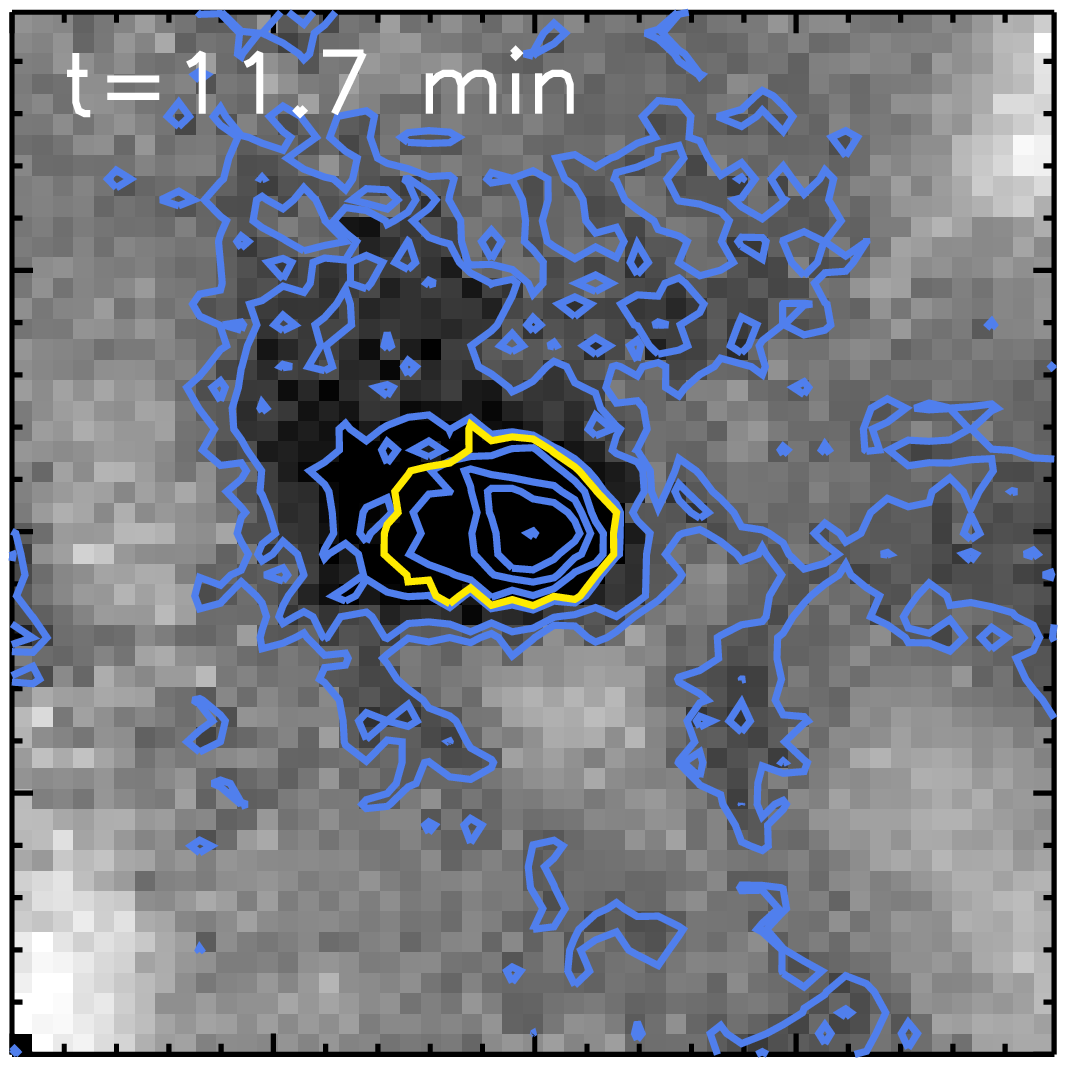}
\hspace{-0.15cm}
\includegraphics[width=0.16\textwidth, bb= 116 400 418 702]{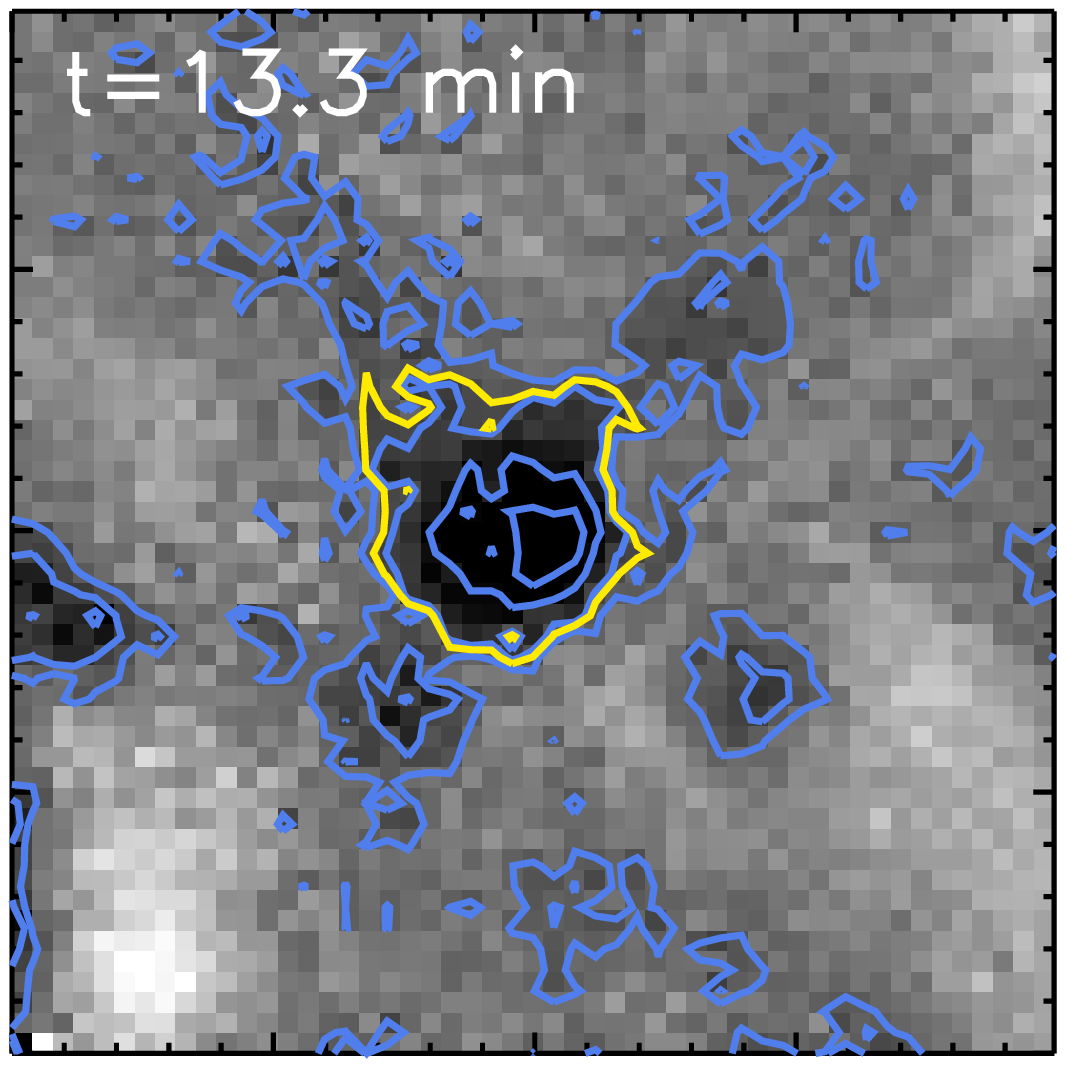}
\hspace{-0.15cm}
\includegraphics[width=0.16\textwidth, bb= 116 400 418 702]{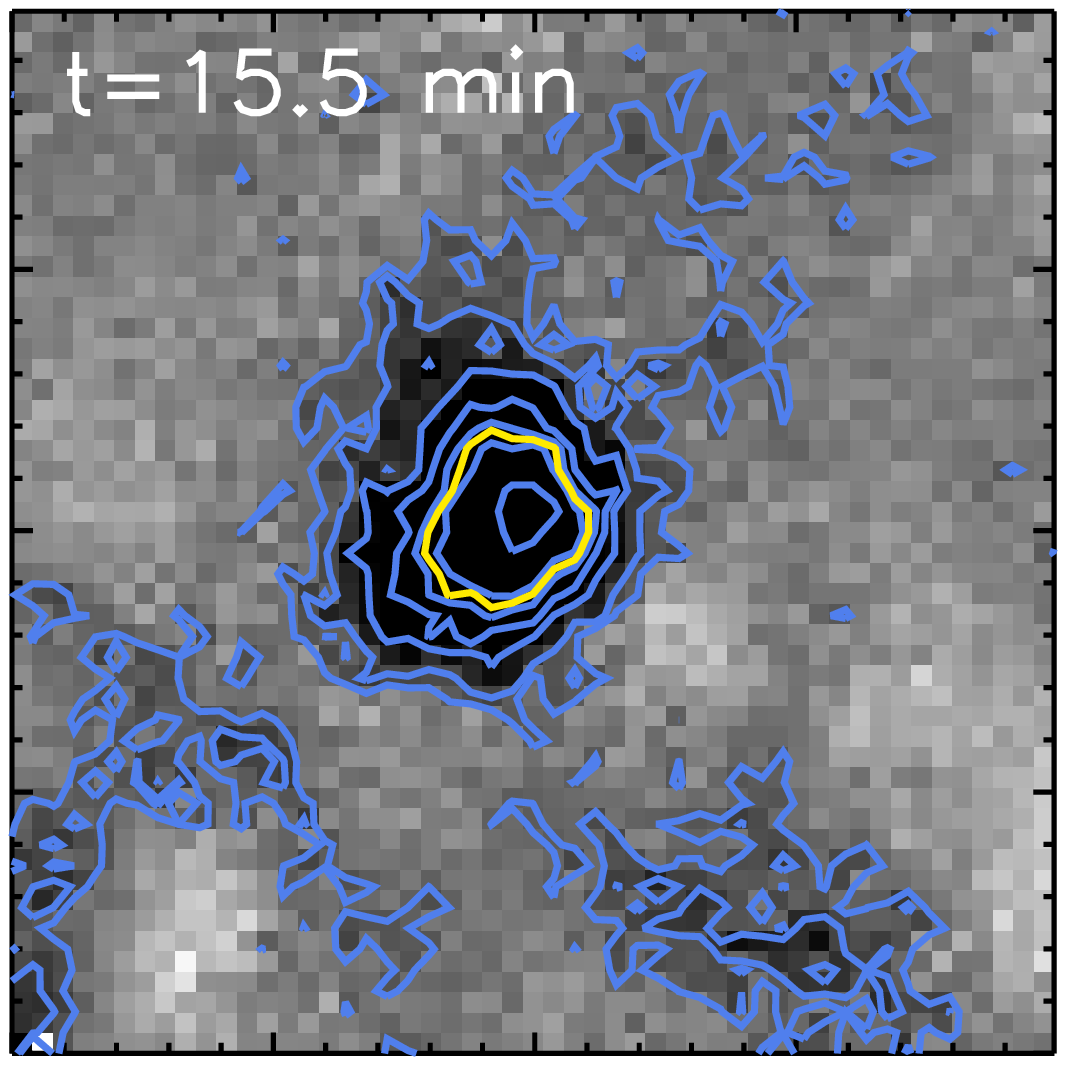}
\hspace{-0.15cm}
\includegraphics[width=0.16\textwidth, bb= 116 400 418 702]{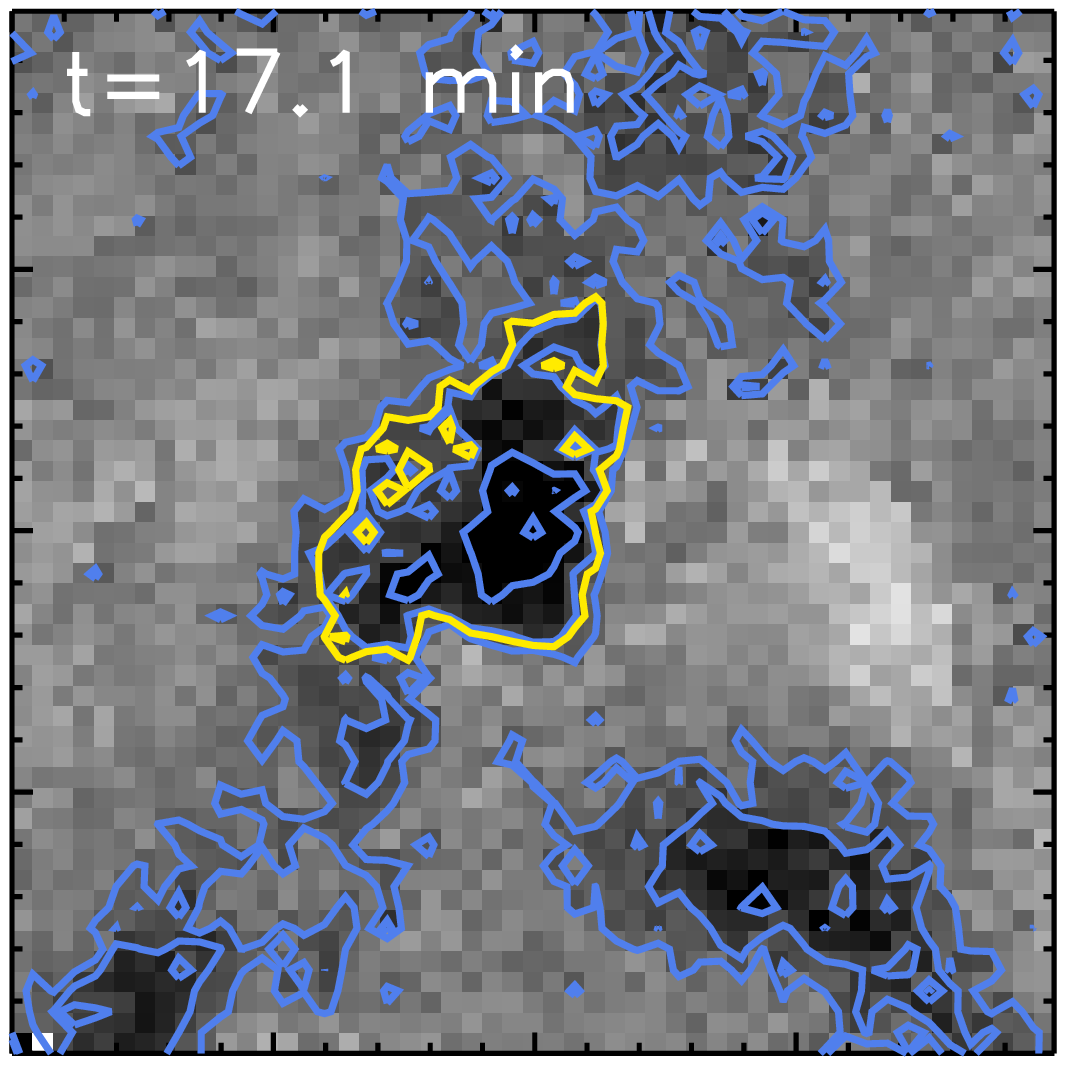}}
\caption{Same as Fig. \ref{ej1} for the top left panel of Fig. \ref{var_area}. Yellow line is the iso-magnetic flux density contour containing a time-constant magnetic flux of -5$\times 10^{16}$ Mx.}
\label{ej2}
\end{figure*}

\section{Oscillations of quiet Sun magnetic fields}

We focus on the temporal evolution of circular polarization patches. Magnetic patches are passively buffeted by granular motions and interact with them, fragmenting, mixing with patches of opposite or equal polarity, and cancelling. This behavior has been also observed at lower spatial resolution \citep[e.g.][]{martin_88}, but at higher resolution we observe more polarization patches and hence a large probability of interaction. In fact, we estimate that roughly 50\% of the circular polarization signals suffers at least one encounter with a neighboring patch in 5 min.

The spatial resolution of the IMaX data ($\sim 0.15''$) allows us to observe an interesting dynamic property of the weak quiet Sun polarization patches. The signals are found to vary in shape, i.e., the area enclosed in a contour containing a constant magnetic flux changes with time. This can be seen in Figs. \ref{ej1} and \ref{ej2}, following the yellow contours. Both the area and the shape clearly change with time.

\begin{figure*}[!t]
\includegraphics[width=0.5\columnwidth]{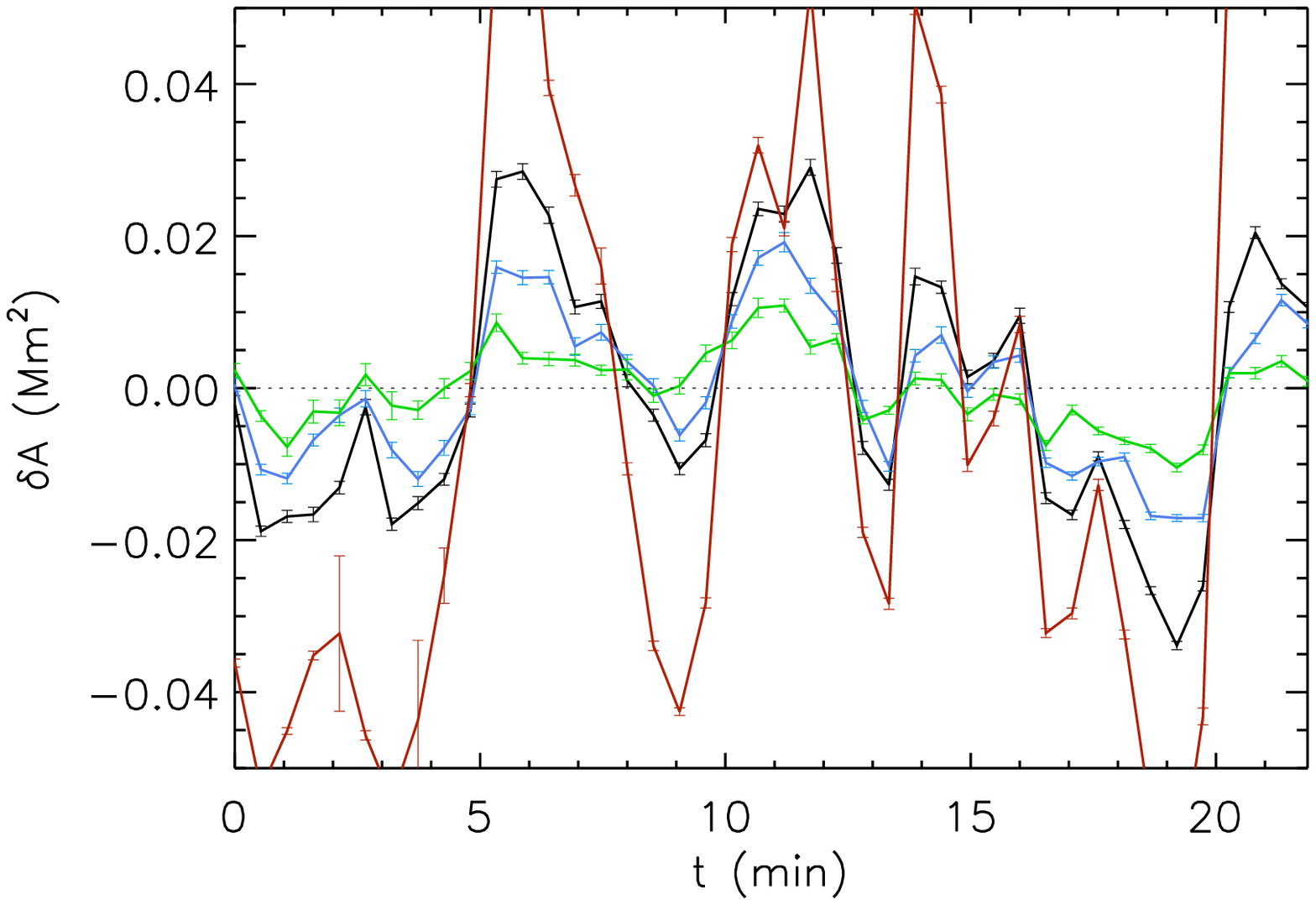}
\includegraphics[width=0.5\columnwidth]{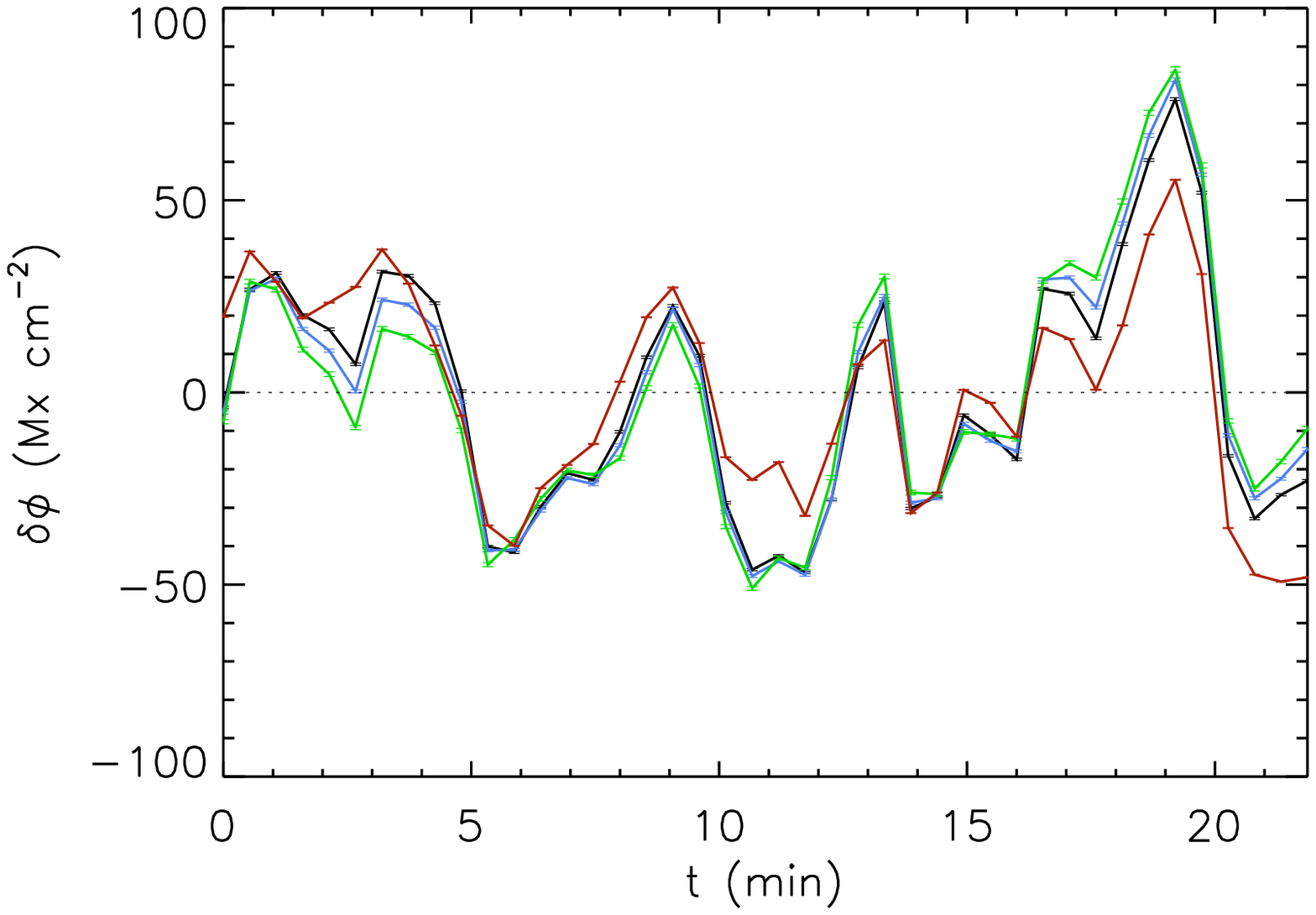}
\caption{Evolution of the area (left) and magnetic flux density (right) of an intense patch (with a maximum value of \hbox{$270$ Mx cm$^{-2}$}). The red line contain 1$\times 10^{18}$ Mx. Note that the oscillation in area of the red line has been divided by 4. The black line represent the fluctuation of the area that contains a constant magnetic flux of 3$\times 10^{17}$ Mx. The blue and green lines are the areas containing 2$\times 10^{17}$ and 1$\times 10^{17}$ Mx, respectively.  The larger the flux (area) the further the contour lies from the center of the structure, which is defined as the position of the maximum flux density.}
\label{network}
\end{figure*}

Although this behavior is common to all magnetic flux density patches (the ones with significant spatial coherency), we select some particular examples to study the area fluctuations in detail. We study four low flux patches ($10^{16}-10^{17}$ Mx) that do not interact with another patch during the whole sequence or for a long time (at least at our present noise level) and an intense patch ($10^{18}$ Mx). We follow the time evolution of these patches at constant magnetic flux. In other words, the patch is defined as the signal enclosed in the magnetic flux density contour for which the magnetic flux is constant with time. At all times, the contour value is above \hbox{8 Mx cm$^{-2}$}, which is twice the noise level of the magnetogram. The areas are enclosed in contours that are treated as polygons, hence, we increase the precision on the computation. Since the magnetic flux density has an associated error, the contours (and the areas) have also an uncertainty. We have estimated the errors associated to the areas numerically using a monte-carlo approach. The larger the area the smaller the relative error. In our case, the errors range between 1-8\%. These uncertainties have been propagated as gaussian variables to derived quantities, such as the mean magnetic flux density enclosed in an area (see Fig. \ref{network}).

\begin{figure*}[!t]
\centering
\includegraphics[width=0.49\textwidth, bb=36 16 479 337]{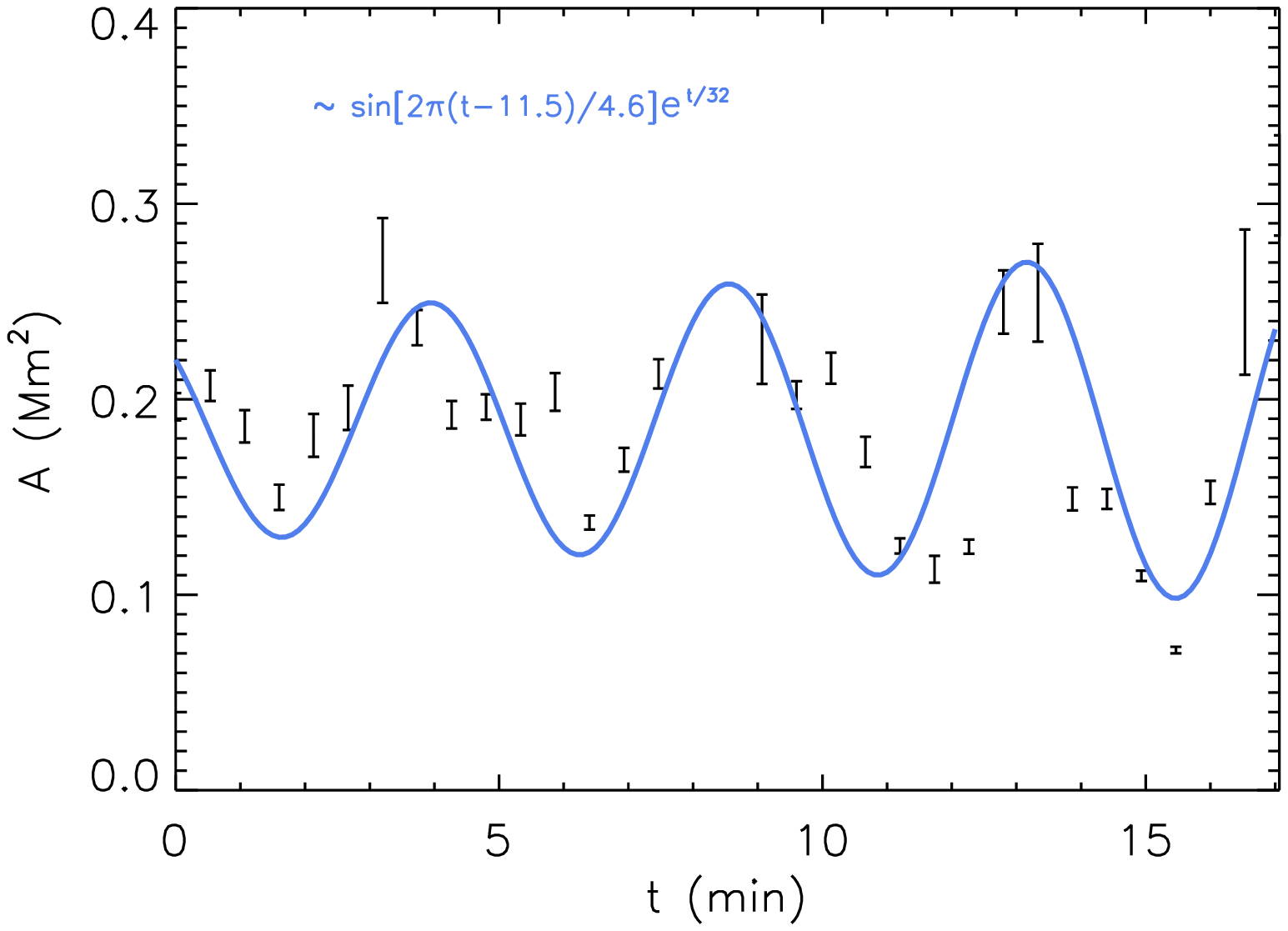}
\includegraphics[width=0.49\textwidth, bb=36 16 479 337]{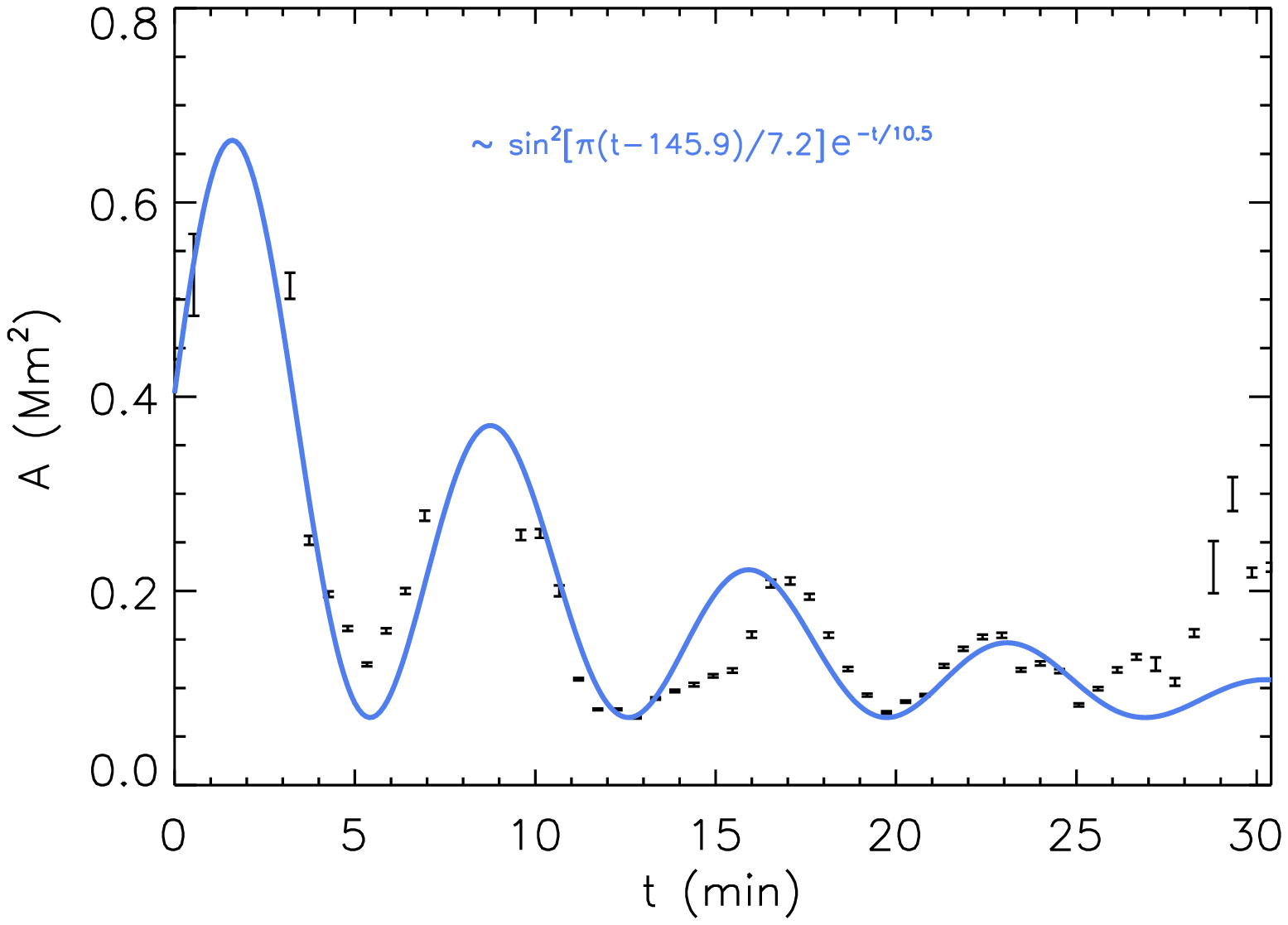}
\includegraphics[width=0.49\textwidth, bb=36 16 479 337]{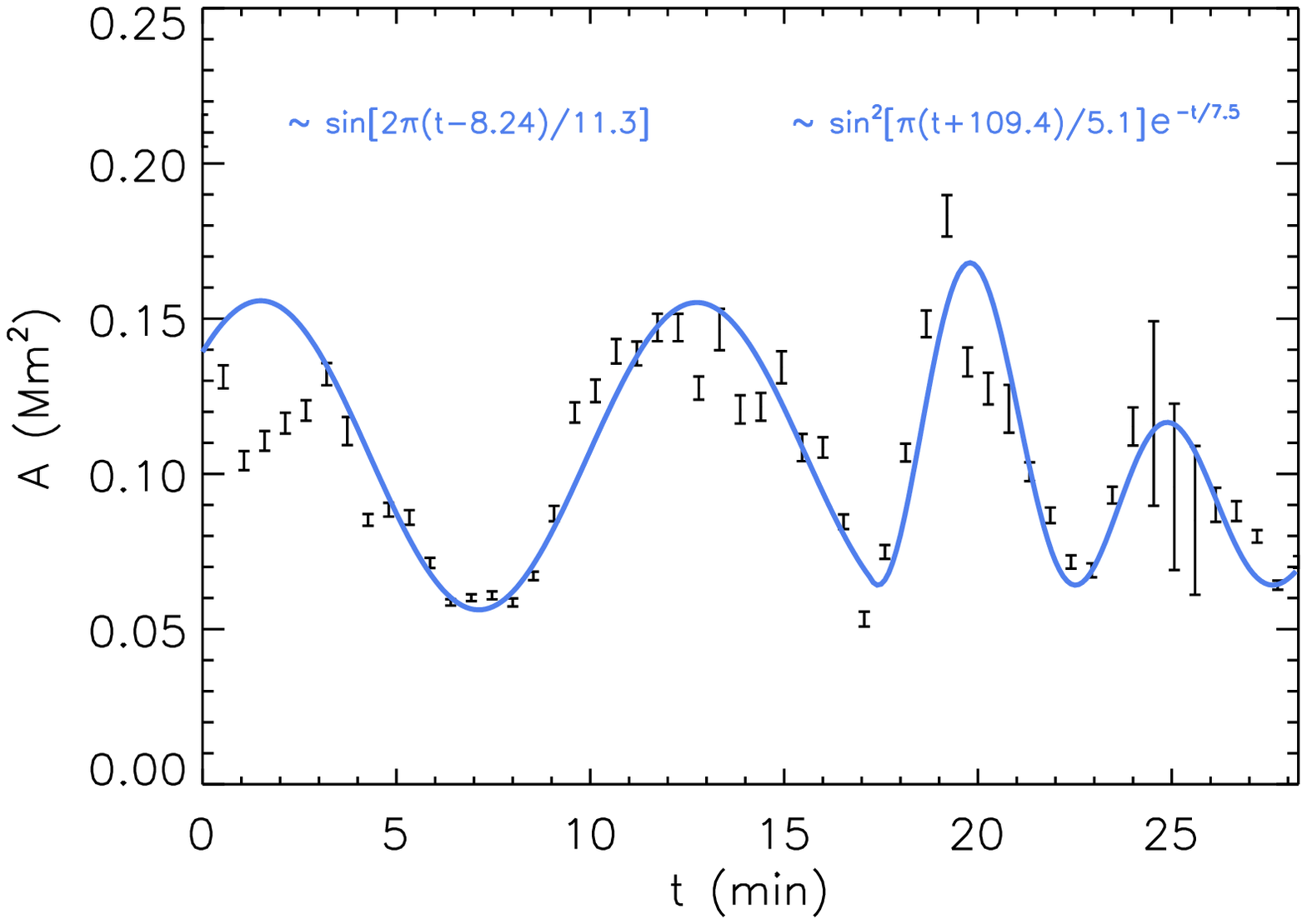}
\includegraphics[width=0.49\textwidth, bb=36 16 479 337]{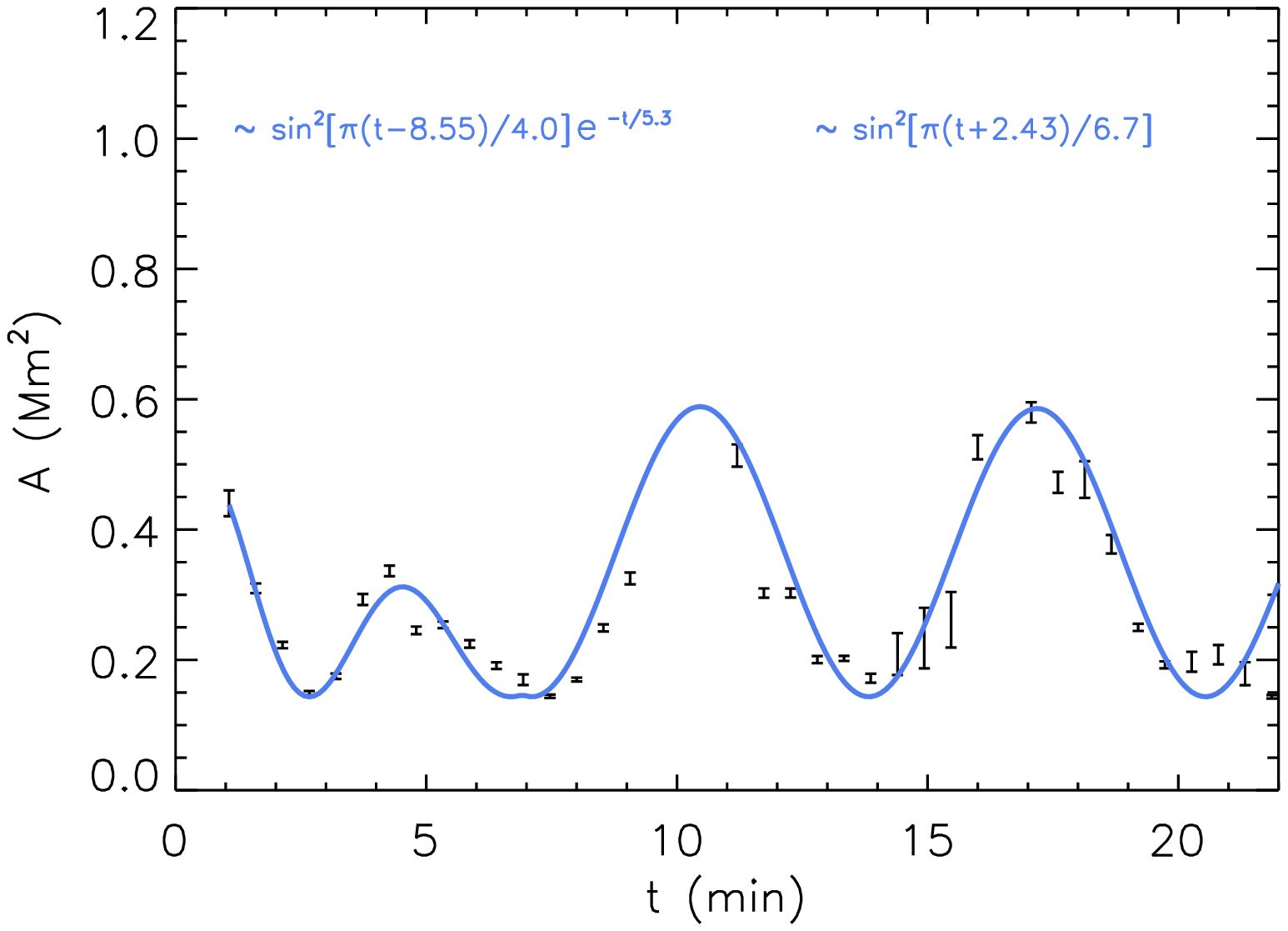}
\label{var_area}
\caption{Evolution of the area of four selected magnetic patches. Constant amounts of magnetic flux are followed in time, being -4.5$\times 10^{16}$, -1.0$\times 10^{17}$, 5$\times 10^{16}$, and -9$\times 10^{16}$ Mx, from the top left panel to the bottom right one. The measured areas are represeted as error bars. The blue lines are analytical fits (using a least-squares estimator) to facilitate recognition of the quasi-periodic pattern. The time $t$ in these expressions is given in minutes.}
\end{figure*}

Figure \ref{var_area} shows the time-dependence of the area of four selected weak magnetic features that show clear periodic fluctuations of the area. The area fluctuates between a factor of 7 and a factor of 2. The four selected cases present a large variety of behaviors. From the top left, clockwise, the modes are amplified, damped and showing two different periods and different damping or amplification factors. The periods range between 4 and 11 minutes, showing that there is not a characteristic period for these oscillatory patterns. These periods have been obtained using analytical fits to the area oscillation since the duration of these events does not allow a reliable fourier analysis. The corresponding anti-phase oscillations of the magnetic flux density have peak to peak variations from 12 up to 100 Mx cm$^{-2}$ around the mean value, which are $-24$, $-67$, 48 and $-34$ Mx cm$^{-2}$ from the top left of Fig. \ref{var_area}, clockwise.

Figure \ref{network} shows the area oscillations of a more intense patch (with a maximum value of \hbox{$270$ Mx cm$^{-2}$}) at three different radii from the center of the structure, defined by the amount of flux contained within the considered perimeter. As for the weaker flux patches, the oscillation contains several periods. In this case, 3 and 5 minutes are the most evident. The area oscillation mantains period and phase over different fluxes. The only difference regards its amplitude, which increases as we move away from the center of the patch. The fluctuation of the area relative to its mean value ranges from 14\%-24\% from the most central to the most peripheral contour, i. e., the areas containing 1$\times 10^{17}$ and 1$\times 10^{18}$ Mx, respectively. The magnitude of the fluctuation of the magnetic flux density remains constant. This means that the variation of this quantity across the structure is very smooth.

\begin{figure*}[!t]
\centering
\includegraphics[width=0.49\textwidth]{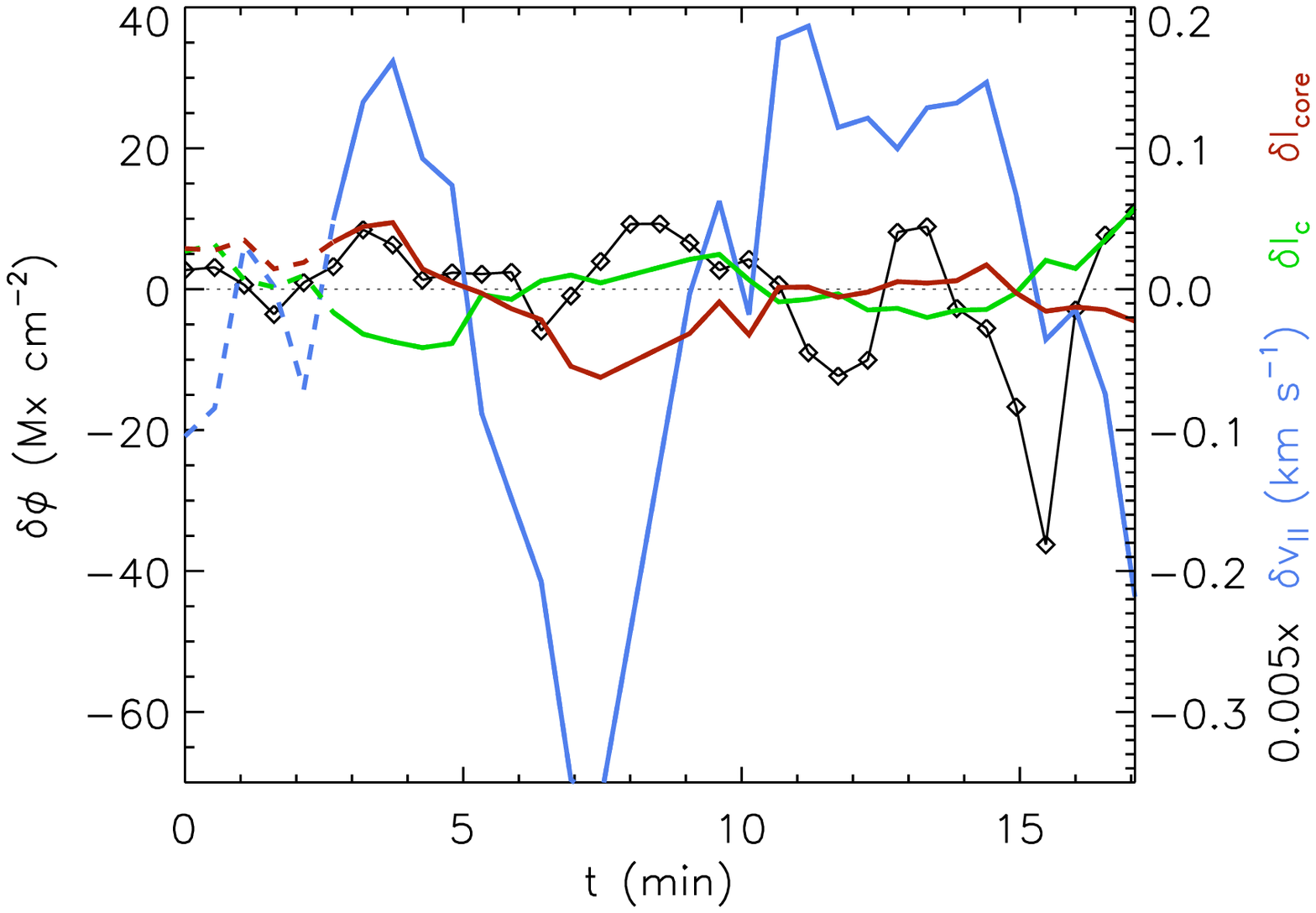}
\includegraphics[width=0.49\textwidth]{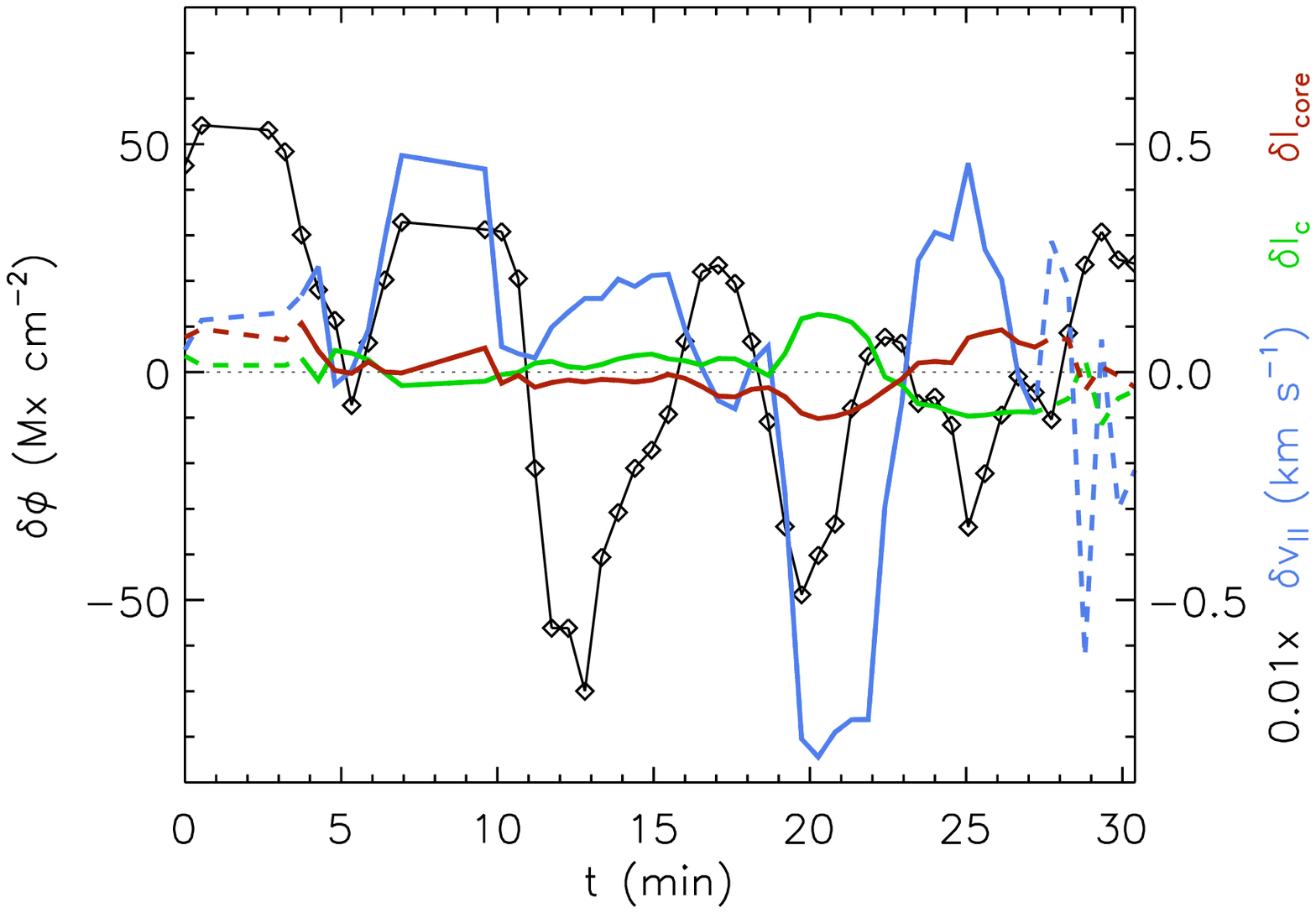}
\includegraphics[width=0.49\textwidth]{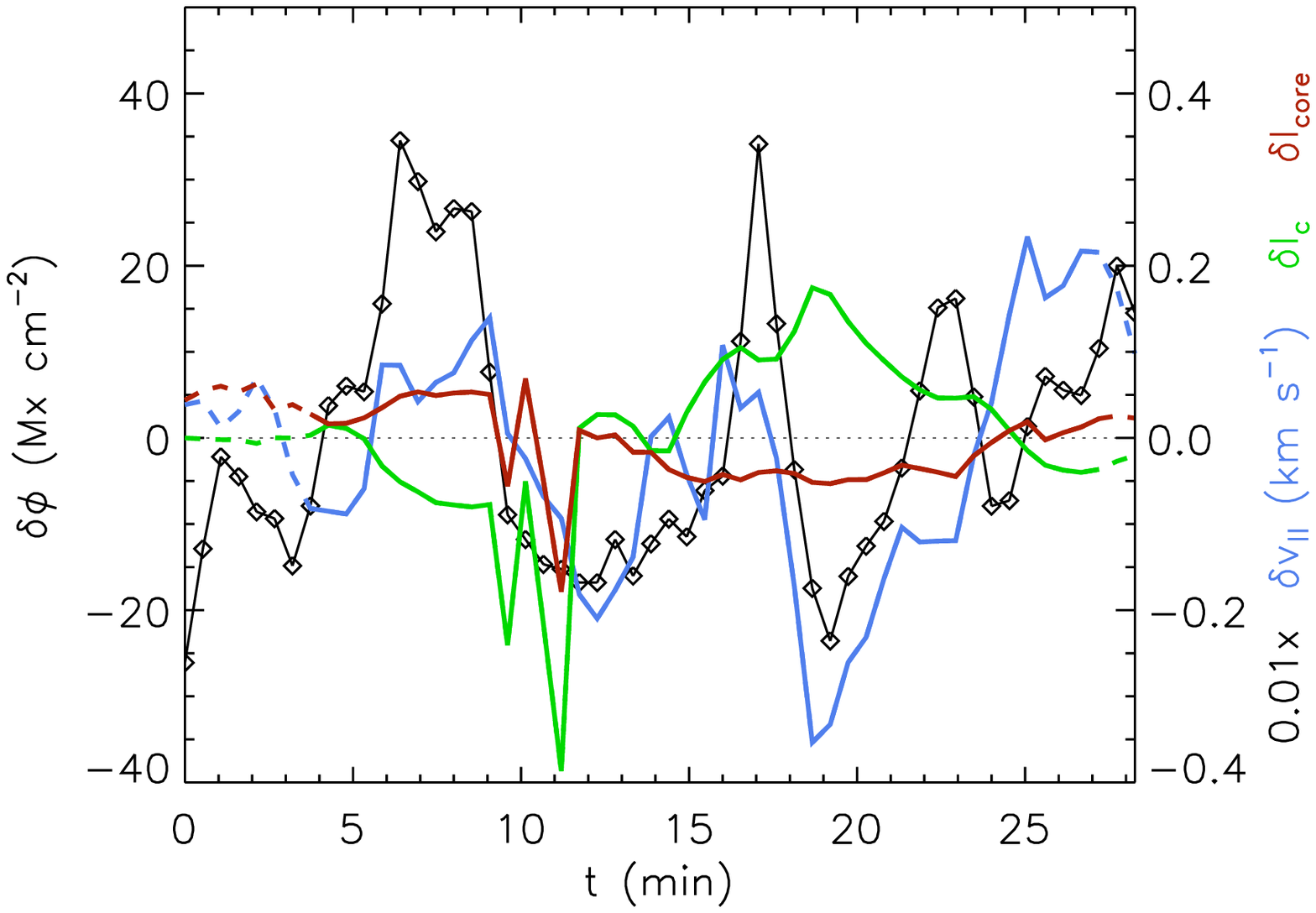}
\includegraphics[width=0.49\textwidth]{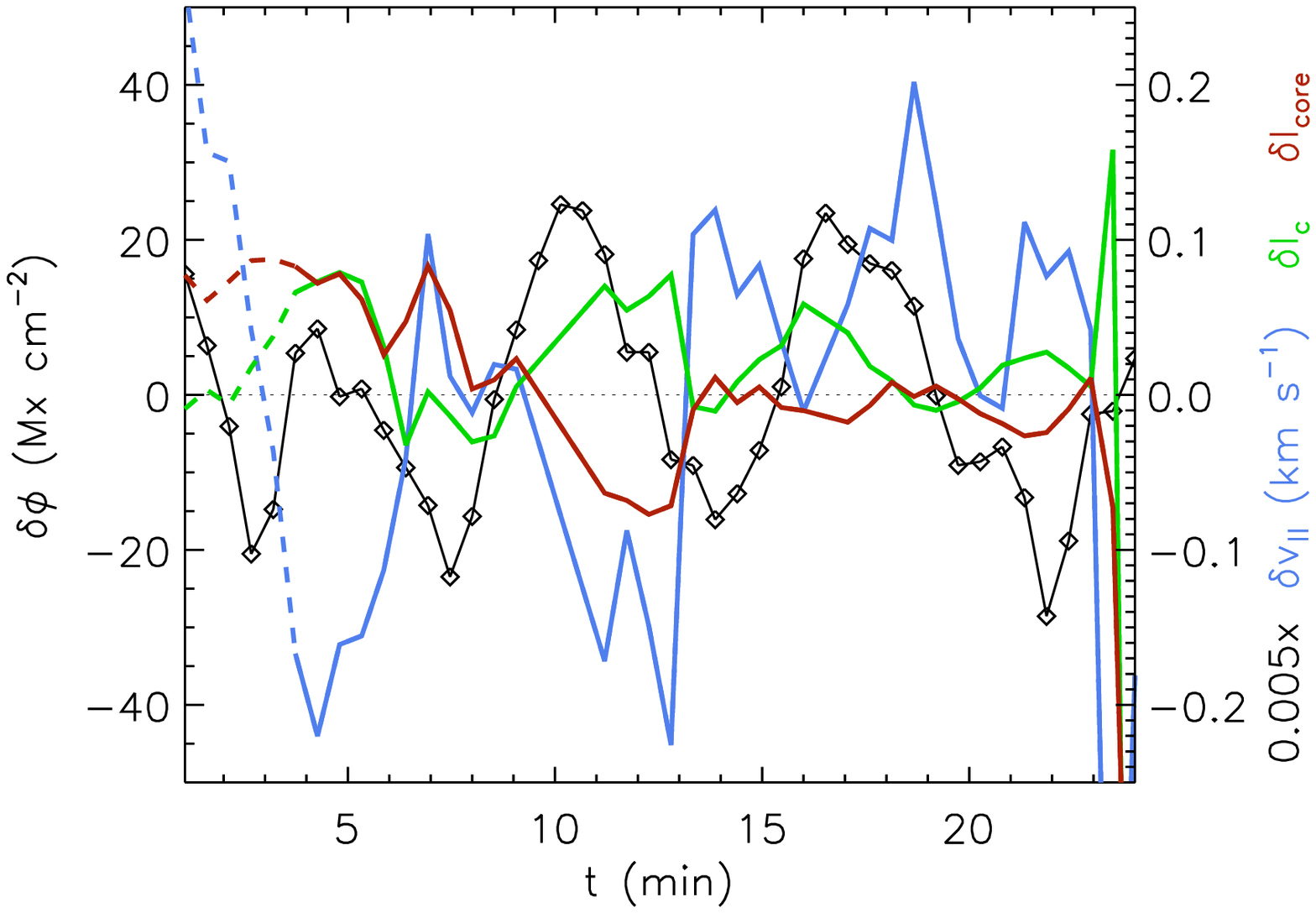}
\label{var_resto}
\caption{Evolution of diverse physical quantities of the four patches of Fig. \ref{var_area}. The black lines represent the oscillation of the magnetic flux density with respect to its time average. Blue, green and red lines represent the variation of the line of sight velocity, continuum and core intensity with respect to its mean value, respectively. Note that these three last quantities have been multiplied by a factor for visualization purposes. Dashed lines show the time interval that is affected by apodisation.}
\end{figure*}

The line of sight velocity, the continuum and the core intensity oscillate too, showing the characteristic pattern of the p-modes, with a 5 minutes period. The continuum and core intensity fluctuations exhibit a phase shift of 180$^\circ$. The velocity is also in anti-phase with the continuum intensity. After removing the effect of p-modes using a subsonic filter, the speed of sound being \hbox{4 km s$^{-1}$}. Neither the oscillation of the velocity, the continuum nor the core intensity seem to be related to the magnetic flux density (and area) fluctuations, as can be seen in Fig. \ref{var_resto}.

Whereas some of the oscillations in area we observe display periods in the range of the p-modes, others have distinctly longer periods, even up to 11 min, and show an abrupt change of the wave period at a given time. Moreover, we find that the area of some magnetic patches located close together oscillate in phase. In one particular case, the co-oscillating flux patches surround a granule, the mean distance between them being about 1600 km.


\section{Discussion}

We have shown that the magnetic field in the quiet Sun oscillates
with a variety of periods compatible with the granular life-time.
These oscillations can be strongly damped (top right panel of Fig.
\ref{var_area}) or amplified (top left panel of Fig.
\ref{var_area}) with a damping time of $\sim 5-30$ min. We have
also observed that the period of the oscillation can change
abruptly in the course of the time series. This would suggest that
either the characteristic modes of the structure vary due to
changes in their physical properties (density, magnetic field
orientation, sub-resolution structure,etc.), or that they are not
characteristic modes of oscillation at all.



Theoretical studies of the characteristic oscillations modes of
thin flux tubes \citep[for example the classical work
by][]{Edwin+Roberts1983}  suggest a dependence of the oscillation
period on the structure size $a$ multiplied by the wave number
$k$, in a sense that structures with larger $ka$ should oscillate
with larger periods. The wave modes excited in flux tubes by
convection can, in principle, have a variety of wave numbers $k$.
However, from the theory of the wave excitation (see for example
\cite{Goldreich+Kumar1990}) one may argue that maximum of the
excitation happens at wavelengths close to the local scale height
in the atmosphere, providing that the modes with the wavelength
about the tube radius (($\sim 250$ km in our case) will have more
power. This argument allows us to fix $k$ and to speculate that,
in principle, based on the work of \cite{Edwin+Roberts1983}, we
can expect that magnetic structures with larger radius $a$ would
oscillate with a larger period.
Nevertheless, the strength of this dependence is difficult to
estimate in our case, as one should know precisely the
thermodynamic and magnetic parameters of the medium inside and
outside the magnetic feature. In any case, the time-average size
of the magnetic features we study in this letter is approximately
the same and, thus, the presence of characteristic oscillation
modes would not explain the scatter in the temporal periods we
observe between the different features. It will also be difficult
to explain the abrupt change in the periods of the oscillation of
a single feature and why some oscillations are damped while others
are amplified. But more importantly, the variations in the area we
detect are so strong that, possibly, a non-linear analysis of the
variations is required, hence, the classical models are not
suitable to explain these observations.

If not being flux tube oscillation modes, the pattern we observe
might be the forcing due to the evolution of the granules. Since
the magnetic fields on the quiet Sun have mainly strengths lower
than the equipartition field in the photosphere \citep[$\sim
300-500$ G; see e.g.][]{lin_95, khomenko_03, marian_08, david_07},
the field lines are continuously buffeted by granular flows, being
squashed or released everywhere and hence the magnetic fields are
constantly being amplified or weakened. However, the present
observations do not allow us to reject either scenario.

Now, an interesting question arises: do these waves propagate up
through the solar atmosphere? We predict that, in case these waves
propagate across the solar atmosphere, the very same phenomena
reported in this letter should be observed in the
filter-polarimeters onboard the Hinode satellite. The different
instruments onboard the Hinode satellite provide valuable
simultaneous data tracing the solar atmosphere \citep[see e.
g.][]{marian_09}. This data may be interesting to future work
devoted to a detailed study of the propagation of these waves from
the photosphere to the chromosphere. Wave propagation is an
efficient means of carrying energy between different atmospheric
heights and of dissipating it efficiently, mainly, through the
formation of shocks. Therefore, we should put additional efforts
into unveiling the role of this newly detected magnetic field
oscillation in heating atmospheric layers.

\begin{acknowledgements}

We are grateful to Manolo Collados, Antonio D\' iaz, and Jaume Terrades for useful suggestions. The Spanish contribution has been funded by the Spanish Ministry of Science and Innovation under projects ESP2006-13030-C06, AYA2009-14105-C06 (including European FEDER funds), and  AYA2010-18029 (Solar Magnetism and Astrophysical Spectropolarimetry). The German contribution to Sunrise is funded by the Bundesministerium f\"ur Wirtschaft und Technologie through Deutsches Zentrum f\"ur Luft- und Raumfahrt e.V. (DLR), Grant No. 50 OU 0401, and by the Innovation fond of the President of the Max Planck Society (MPG). This work has been partially supported by the WCU grant No. R31-10016 funded by the Korean Ministry of Education, Science, and Technology.

\end{acknowledgements}


\begin{thebibliography}{34}
\expandafter\ifx\csname natexlab\endcsname\relax\def\natexlab#1{#1}\fi

\bibitem[{{Barthol} {et~al.}(2010)}]{barthol_10}
{Barthol}, P., {et~al.} 2010, SoPh, in press

\bibitem[{{Bloomfield} {et~al.}(2006)}]{Bloomfield+etal2006}
{Bloomfield},~{D.S.}, {McAteer},~{R.T.J.}, {Mathioudakis},~{M.},
{Keenan},~{F.P.}, ApJ, 652, 812

\bibitem[{{Centeno} {et~al.}(2009){Centeno}, {Collados}, \& {Trujillo
  Bueno}}]{Centeno+etal2009}
{Centeno}, R., {Collados}, M., \& {Trujillo Bueno}, J. 2009, ApJ,
692, 1211

\bibitem[{{Centeno} {et~al.}(2007)}]{rebe_07}
{Centeno}, R., {et~al.} 2007, ApJ, 666, 137L

\bibitem[{{Danilovic} {et~al.}(2010)}]{danilovic_10}
{Danilovic}, S., {et~al.} 2010, ApJ, 723, L149

\bibitem[{{De Moortel} {et~al.}(2002)}]{DeMoortel+etal2002}
{De Moortel},~{I.}, {Ireland},~{J.}, {Hood},~{A.W.},
{Walsh},~{R.W.}, 2002, A\&A, 387, L13

\bibitem[{{De Pontieu} {et~al.}(2003)}]{DePontieu+etal2003}
{De Pontieu},~{B.}, {Erdelyi},~{R.}, {de~Wijn},~{A.G.}, ApJ, 595,
L63

\bibitem[{{Edwin} \& {Roberts}(1983)}]{Edwin+Roberts1983}
{Edwin}, P. M., \& {Roberts}, B. 1983, Solar. Phys., 88, 179

\bibitem[{{Fujimura} \& {Tsuneta}(2009)}]{fujimura_09}
{Fujimura}, D., \& {Tsuneta}, S. 2009, ApJ, 702, 1443

\bibitem[{{G\"om\"ory} {et~al.}(2010){G\"om\"ory}, {Beck}, {Balthasar},
  {Ryb\'ak}, {Kucera}, {Koza}, \& {W\" ohl}}]{gomory_10}
{G\"om\"ory}, P., {Beck}, C., {Balthasar}, H., {Ryb\'ak}, J., {Kucera}, A.,
  {Koza}, J., \& {W\" ohl}, H. 2010, A\&A, 511, 14

\bibitem[{{Goldreich} \& {Kumar}(1990){Goldreich} \& {Kumar}}]{Goldreich+Kumar1990}
{Goldreich}, P. \& {Kumar}, P. 1990, ApJ, 363, 694

\bibitem[{{Harvey} {et~al.}(2007){Harvey}, {Branston}, \& {Keller}}]{harvey_07}
{Harvey}, J.~W., {Branston}, C.~J., \& {Keller}, C.~U. 2007, ApJ, 177, L180

\bibitem[{{Jess} {et~al.}(2009){Jess}, {Mathioudakis}, {Erd{\'e}lyi},
  {Crockett}, {Keenan}, \& {Christian}}]{jess_09}
{Jess}, D.~B., {Mathioudakis}, M., {Erd{\'e}lyi}, R., {Crockett}, P.~J.,
  {Keenan}, F.~P., \& {Christian}, D.~J. 2009, Science, 323, 1582

\bibitem[{{Khomenko} {et~al.}(2003){Khomenko}, {Collados}, {Solanki}, {Lagg},
  \& {Trujillo Bueno}}]{khomenko_03}
{Khomenko}, E.~V., {Collados}, M., {Solanki}, S.~K., {Lagg}, A., \& {Trujillo
  Bueno}, J. 2003, A\&A, 408, 1115

\bibitem[{{Krijger} {et~al.}(2001)}]{Krijer+etal2001}
{Krijger},~{J.M.}, {Rutten},~{R.J.}, {Lites},~{B.W.},
{Straus},~{T.}, {Shine},~{R.A.}, {Tarbell},~{T.D.}, 2001, A\&A,
379, 1052

\bibitem[{{Lagg} {et~al.}(2010)}]{lagg_10}
{Lagg}, A., {et~al.} 2010, A\&A, 723, L164

\bibitem[{{Lin}(1995)}]{lin_95}
{Lin}, H. 1995, ApJ, 446, 421

\bibitem[{{Lin} \& {Rimmele}(1999)}]{lin_99}
{Lin}, H., \& {Rimmele}, T. 1999, ApJ, 514, 448

\bibitem[{{Lites} {et~al.}(2008)}]{lites_08}
{Lites}, B.~W., {et~al.} 2008, ApJ, 672, 1237

\bibitem[{{Lites} {et~al.}(1993)}]{Lites+Rutten+Kalkofen1993}
{Lites},~{B.W.}, {Rutten},~{R.J.}, {Kalkofen},~{W.} 1993, ApJ,
414, 345

\bibitem[{{Manso Sainz} {et~al.}(2010){Manso Sainz}, {Mart\' inez Gonz\'alez},
  \& {Asensio Ramos}}]{rafa_10}
{Manso Sainz}, R., {Mart\' inez Gonz\'alez}, M.~J., \& {Asensio Ramos}, A.
  2010, ApJ, submitted

\bibitem[{{Mart\' inez Gonz\'alez} {et~al.}(2011){Mart\' inez Gonz\'alez}, {Manso Sainz},
  \& {Asensio Ramos}}]{marian_11}
{Mart\' inez Gonz\'alez}, M.~J., {Manso Sainz}, R., \& {Asensio Ramos}, A.
  2011, MNRAS, in preparation

\bibitem[{{Mart\' inez Gonz\'alez} \& {Bellot Rubio}(2009)}]{marian_09}
{Mart\' inez Gonz\'alez}, M.~J., \& {Bellot Rubio}, L. 2009, ApJ, 700, 1391

\bibitem[{{Mart\' inez Gonz\'alez} {et~al.}(2008){Mart\' inez Gonz\'alez},
  {Collados}, {Ruiz Cobo}, \& {Beck}}]{marian_08}
{Mart\' inez Gonz\'alez}, M.~J., {Collados}, M., {Ruiz Cobo}, B., \& {Beck}, C.
  2008, A\&A, 477, 953

\bibitem[{{Mart\' inez Gonz\'alez} {et~al.}(2007){Mart\' inez Gonz\'alez},
  {Collados}, {Ruiz Cobo}, \& {Solanki}}]{marian_07}
{Mart\' inez Gonz\'alez}, M.~J., {Collados}, M., {Ruiz Cobo}, B., \& {Solanki},
  S.~K. 2007, A\&A, 469, 39

\bibitem[{{Mart\' inez Gonz\'alez} {et~al.}(2010){Mart\' inez Gonz\'alez},
  {Manso Sainz}, {Asensio Ramos}, \& {Bellot Rubio}}]{marian_10}
{Mart\' inez Gonz\'alez}, M.~J., {Manso Sainz}, R., {Asensio Ramos}, A., \&
  {Bellot Rubio}, L. 2010, ApJ, 714L, 94

\bibitem[{{Mart\' inez Pillet} {et~al.}(2010)}]{valentin_10}
{Mart\' inez Pillet}, V., {et~al.} 2010, SoPh, in press

\bibitem[{{Martin}(1988)}]{martin_88}
{Martin}, S.~F. 1988, SoPh, 117, 243

\bibitem[{{Orozco Su\'arez} {et~al.}(2008){Orozco Su\'arez}, {Bellot Rubio},
  {del Toro Iniesta}, \& {Tsuneta}}]{david_08}
{Orozco Su\'arez}, D., {Bellot Rubio}, L.~R., {del Toro Iniesta}, J.~C., \&
  {Tsuneta}, S. 2008, A\&A, 481, 330

\bibitem[{{Orozco Su\'arez} {et~al.}(2007)}]{david_07}
{Orozco Su\'arez}, D., {et~al.} 2007, ApJ, 670, 61

\bibitem[{{Solanki} {et~al.}(2010)}]{sami_10}
{Solanki}, S.~K., {et~al.} 2010, ApJ, 723, L127

\bibitem[{{Stenflo} {et~al.}(1987){Stenflo}, {Solanki}, \&
  {Harvey}}]{stenflo_solanki_87}
{Stenflo}, J.~O., {Solanki}, S.~K., \& {Harvey}, J.~W. 1987, A\&A, 171, 305

\bibitem[{{Vecchio} {et~al.}(2009){Vecchio}, {Cauzzi}, \& {Reardon}}]{Veccio+etal2009}
{Vecchio}, A., {Cauzzi}, G., \& {Reardon}, K. P. 2009, A\&A, 494,
269

\bibitem[{{Wiegelmann} {et~al.}(2010)}]{wiegelmann_10}
{Wiegelmann}, T., {et~al.} 2010, ApJ, 723, L185

\end{thebibliography}

\end{document}